\documentclass[11pt,a4paper]{article}
\pdfoutput=1   % required for arXiv submission if pdflatex is used
\usepackage{amsmath,amsfonts,amssymb,amsbsy}
\usepackage{mathrsfs,latexsym}
\usepackage{graphicx}
\usepackage{color}
\usepackage{booktabs}
\usepackage{multirow}
\usepackage{multicol}
\usepackage{subfig}
\usepackage{alpha}
\usepackage{mathrsfs}
\hypersetup{%
  pdftitle = {The decoupling of heavy sea quarks},
  pdfauthor = {M.Athenodorou, J.Finkenrath, F.Knechtli, T.Korzec, B.Leder, M.Marinkovic, R.Sommer},
  pdfkeywords = {Lattice QCD, Decoupling, Effective theory, Matching of Lambda parameters,  Charm quark}
}%
\usepackage{macros_alpha}
\usepackage{textcomp}

\newcommand{\Lameff}{\Lambda_\tl}
\newcommand{\gbareff}{\overline{g}_\tl}
\newcommand{\gbarnl}{\overline{g}_{\tl}}
\newcommand{\gbarnq}{\overline{g}_{\tq}}
\newcommand{\betanl}{\beta_{\tl}}
\newcommand{\betanq}{\beta_{\tq}}
\renewcommand{\bzero}{b_0(\nq)}
\newcommand{\bzerop}{b_0(\nl)}

\newcommand{\br}{\tilde{b}}
\newcommand{\Cs}{\widetilde{C}}
\newcommand{\logML}{L_\mathrm{M}}

\newcommand{\nq}{N_\mathrm{f}}
\newcommand{\tq}{\mathrm{f}}
\newcommand{\tl}{\mathrm{\ell}}
\newcommand{\Lamq}{\Lambda_{\tq}}
\newcommand{\Laml}{\Lambda_{\tl}}
\newcommand{\mscale}{{\cal S}}

\newcommand{\ev}[1]{\left\langle #1 \right\rangle}

\begin{document}

\preprintno{%
DESY 18-134
\\
HU-EP-18/28
\\
WUB/18-03
}

\title{How perturbative are heavy sea quarks?
}

\collaboration{\includegraphics[width=2.8cm]{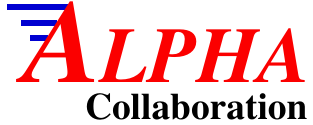}}

\author[cyp,cyi]{Andreas~Athenodorou}
\author[cyi]{Jacob~Finkenrath}
\author[wup]{Francesco~Knechtli}
\author[wup]{Tomasz~Korzec}
\author[hu]{Bj{\"o}rn~Leder}
\author[trin]{Marina~Krsti{\'c}~Marinkovi{\'c}}
\author[desy,hu]{Rainer~Sommer}

\address[cyp]{Department of Physics, University of Cyprus,
              P.O. Box 20537, Nicosia CY, Cyprus}
\address[cyi]{CaSToRC, CyI Athalassa Campus,
              20~Constantinou Kavafi Street, 2121~Nicosia, Cyprus}
\address[wup]{Department of Physics, Bergische Universit{\"a}t Wuppertal, 
              Gaussstr.~20, 42119~Wuppertal, Germany}
\address[hu]{Institut~f\"ur~Physik, Humboldt-Universit\"at~zu~Berlin,
             Newtonstr.~15, 12489~Berlin, Germany}
\address[trin]{School~of~Mathematics, Trinity~College, Dublin~2, Ireland}
\address[desy]{John von Neumann Institute for Computing (NIC),
               DESY, Platanenallee~6, D-15738 Zeuthen, Germany}

\vspace*{-1cm}

\begin{abstract}
Effects of heavy sea quarks on the low energy physics are
described by an effective theory where the expansion parameter is the inverse 
quark mass, $1/M$. At leading order in $1/M$ (and neglecting
light quark masses) the dependence of any low energy quantity on the
heavy quark mass is given in terms of the ratio of $\Lambda$ parameters
of the effective and the fundamental theory. 
We define a function describing the scaling with the mass $M$.
Our study of perturbation theory  suggests
that its perturbative expansion is very reliable for the bottom quark and also seems to work very well at the charm quark mass.
The same is then true for the ratios of $\Lambda^{(4)}/\Lambda^{(5)}$ and $\Lambda^{(3)}/\Lambda^{(4)}$, which play
a major r\^ole in connecting (almost all) lattice determinations
of $\alpha^{(3)}_\msbar$ from the three-flavor theory with 
$\alpha^{(5)}_\msbar(M_\mathrm{Z})$.
Also the charm quark content of the nucleon, relevant for dark matter searches, can be computed accurately from perturbation theory.

In order to further test perturbation theory in this situation,
we investigate a very closely related model, namely QCD with 
$\nq=2$ heavy quarks. Our non-perturbative information is
derived from simulations on the lattice,
with masses up to the charm quark mass and lattice spacings 
down to about 0.023 $\fm$ followed by a continuum extrapolation.
The non-perturbative mass 
dependence agrees within rather small errors with the 
perturbative prediction at masses around the
charm quark mass. Surprisingly,  
from studying solely the massive theory we can make 
a prediction for the ratio $Q^{1/\sqrt{t_0}}_{0,2}=
[\Lambda \sqrt{t_0(0)}]_{\nq=2} / [\Lambda \sqrt{t_0}]_{\nq=0}$,
which refers to the chiral limit in $\nq=2$.
Here $t_0$ is the Gradient Flow scale of \cite{flow:ML}.
The uncertainty for $Q$ is estimated 
to be 2.5\%. For the phenomenologically interesting 
$\Lambda^{(3)}/\Lambda^{(4)}$, we conclude that perturbation
theory introduces errors which are at most at the 1.5\% level,
smaller than other current uncertainties.

\end{abstract}


\maketitle

\tableofcontents

\section{Introduction}

At present most simulations of lattice Quantum Chromodynamics (QCD) include
two light (up and down) quarks and a strange quark. It is important to
investigate the effects of the charm quark, whose mass $M$ is about 12 times
larger than that of the strange quark. 
Effective field theory \cite{Weinberg:1980wa}
arguments predict that the effects of a heavy
quark are described by the theory without the heavy quark with leading
order power corrections of size O$(1/M^2)$.
At lowest order in $1/M$  only the light quark masses
and the coupling need to be adjusted to
match the two theories (with and without the heavy quarks). 
For the coupling this
issue has been discussed in perturbation theory in \cite{thresh:BeWe}.   
The matching of the coupling in the case of the decoupling of one heavy quark
is known to four loops in perturbation theory 
\cite{Chetyrkin:2005ia,Schroder:2005hy}.
Equivalently to match the couplings at a given renormalization scale
one can formulate a relation between the
renormalization group invariant $\Lambda$ parameters of
the effective and the fundamental theory.
In this article we present a study of the perturbative behaviour of the ratio of
$\Lambda$ parameters computed up to four loops in the matching of the
couplings, which requires the knowledge of the five loop $\beta$ function,
which had been computed in Refs.~\cite{vanRitbergen:1997va,Czakon:2004bu,Baikov:2016tgj,Luthe:2016ima,Herzog:2017ohr}.

Besides studying the behaviour of perturbation theory itself it is desirable to
compare to non-perturbative data. This is especially the case for the charm
quark given that matching is performed at a fairly low scale $\approx1.3\, \GeV$
in this case. 
It is very difficult to compare directly
2+1 flavor and 2+1+1 flavor lattice simulations, because various systematic
uncertainties mask the physical effect.
We proposed instead to simulate a model,
namely QCD with two heavy, mass-degenerate quarks \cite{Bruno:2014ufa}.
The effective theory is the Yang-Mills theory up to $1/M^2$ corrections.
The mass dependence of ratios of hadronic scales such as
$\sqrt{t_0(0)}/\sqrt{t_0(M)}$, where $t_0$ is the Gradient flow 
scale \cite{flow:ML}
factorizes \cite{Bruno:2014ufa} at leading order in a 
non-perturbative and mass-independent factor, and a factor $P$, which is the 
ratio of the $\Lambda$ parameters and depends on the heavy quark mass
through the matching. Since the latter can be evaluated in perturbation
theory we can compare the perturbative 
mass dependence of hadronic scales to the
non-perturbative results from the simulations.
We define a mass-scaling function which is the
logarithmic derivative of $P$ with respect to the logarithm of the
mass. It can be determined directly from the simulations and compared
to its perturbative expansion.

This article is organized as follows. In section \ref{s:decqcd} we describe
the effective theory of decoupling. Section \ref{s:leading} contains a review
of the matching of the effective and fundamental theory at leading order.
We present a perturbative study of the ratio $P$ of the $\Lambda$ parameters,
which results from the matching of the theories at leading 
order, and of the mass-scaling function.
In section \ref{s:simulation} we explain our non-perturbative
study of decoupling in a theory with $\Nf=2$ mass-degenerate heavy fermions
with masses ranging up to (slightly above) the charm quark mass.
We introduce the hadronic scales which we calculate in Monte Carlo simulations
of lattice QCD and give
details of the lattice simulations.
The comparison of the non-perturbative mass dependence of
hadronic scales computed from the simulations with perturbation theory is
presented in section \ref{s:np}.
The implications of these results for the applicability of perturbation theory
at the scale of the charm quark mass are discussed in section \ref{s:disc}.
We summarize our results in section \ref{s:concl}.
In the appendix \ref{s:coefficients} we reproduce
the explicit formulae for the matching of the couplings up to four loops
and the perturbative coefficients of the mass-scaling function.
The asymptotic behavior for large masses of $P$ is derived in
appendix \ref{s:asymptotic}.
Finally appendix \ref{s:tables} contains tables listing the simulations parameters.

\section{The effective theory: decQCD}  \label{s:decqcd}

The effective theory associated with the decoupling of heavy quarks is formally
obtained by integrating out the heavy quark fields. The resulting effective theory
contains a tower of non-renormalizable interactions, which however are suppressed at low
energies by negative powers of the heavy quark masses \cite{Weinberg:1980wa}. The (infinite number of) couplings
of the effective theory can be matched order by order and used to describe the effect of heavy quarks at low energies.

To be precise, let us consider ${\rm QCD}_{\nq}$ with $\nq$ quarks in total, of which $\nl$ are light and $\nq-\nl$ are heavy. For simplicity we assume the light and the heavy quarks to be mass degenerate with the heavy mass given by $M$. Non-degenerate quark masses are conceptually similar, see note at the end of this section. In general the Lagrangian of the effective theory is
\bes
\lag{dec}= \lag{0} + \frac1{M}\lag{1} + \frac1{M^{2}}\lag{2} + \dots\,,
\ees
where the leading order equals ${\rm QCD}_{\nl}$ with $\nl$ quarks and the corrections $\lag{\mathit k}$, $k\ge 1$ consist of linear combinations of local operators of dimension $4+k$. These operators are composed of only the light quark and gauge fields, and include possible light mass factors. They have to satisfy the symmetries of ${\rm QCD}_{\nq}$, most prominently gauge, Euclidean (or Lorentz) and chiral invariance. 
For the  cases of interest, 
 operators of dimension five are excluded and corrections to the leading order start at $\rmO(M^{-2})$
\bes
\label{e:effectiveLag}
\lag{dec}= \lag{QCD_\mathit{\nl}}
+ \frac1{M^{2}} \sum_i \omega_i \Phi_i + \dots\,.
\ees
Here we write $\lag{2}$ explicitly as a linear combination of local dimension six operators $\Phi_i$, multiplied by dimensionless couplings $\omega_i$. 

The simplest situation in which \eqref{e:effectiveLag} holds is $\nl=0$, i.e., when light quarks are absent: the leading order is Yang-Mills theory and there is no gauge invariant dimension five operator made up of gauge fields alone. Thus at leading order only the gauge coupling has to be matched.
We are basing our non-pertubative investigations in sections \ref{s:simulation}-\ref{s:np} on this setting.

In the presence of $\nl\ge 2$ mass-less quarks the non-singlet, non-anomalous chiral symmetry in the light quark sector 
forbids any dimension five operator. The gauge coupling is still the only coupling to be matched at leading order. 
Note that the dynamical (non-perturbative)
breaking of chiral symmetry plays no role here as we may consider 
full and effective theory in a finite (but large) volume where
dynamical symmetry breaking is absent, in full analogy with the elegant derivation of automatic $\rmO(a)$ improvement of
twisted mass QCD in
\cite{Nara:stefan}.
More explicitly
consider a chirally non-invariant observable in the full theory
in finite volume. It vanishes, while a priori in the effective theory at dimension
five the Pauli term $\omega_{\mathrm{Pauli}}\bar{\psi}i\sigma_{\mu\nu}F_{\mu\nu}\psi / M $ contributes as the only dimension five
gauge invariant operator. Matching of full and effective theory requires $\omega_{\mathrm{Pauli}}=0$.

In section \ref{s:leading} we consider the
leading order in $1/M$ in perturbation theory for 
various values of $\nl,\nq$.

For finite light quark masses there are 
dimension five operators, which are formed of the 
operators in $\lag{QCD_\mathit{\nl}}$ multiplied by 
the light quark masses. Their effect can be absorbed in a 
redefinition of the gauge coupling and light quark masses
at the order $m_\mathrm{l} / M$.
The Pauli term multiplied by the light quark masses contributes at dimension six. It is one of the $\Phi_i$ in \eq{e:effectiveLag}. 
Besides the gauge coupling now also the light quark mass needs to be matched. 

All in all, finite light quark masses do not
change the structure of \eq{e:effectiveLag}. Of course couplings in the effective Lagrangian now also depend on the light quark mass. 
The only restriction is that when light quarks
are present, we need at least a doublet, such that there is
a non-anomalous chiral symmetry of the mass-less theory and we can
conclude $\omega_{\mathrm{Pauli}}=0$ as sketched above.

In the following we concentrate on $\nl\ge 2$ mass-less or $\nl=0$ quarks.

\section{Mass-dependence in the leading order effective theory}  \label{s:leading}

At leading order, the only parameter of the
effective theory, ${\rm QCD}_{\nl}$, is its running coupling and
the theory predicts all observables when the coupling is
prescribed at a given renormalization scale in a given renormalization
scheme. It is conceptually cleaner, but completely equivalent in terms
of the physical content to specify the renormalization group invariant (RGI)
$\Lambda$-parameter. The scale dependence of the input is then gone
and the scheme-dependence is easily computable: the one-loop relation
of couplings yields the exact relation of the associated $\Lambda$-parameters.

Explicitly the $\Lambda$-parameter of QCD with $\nf$ quarks,
 \bes
 \Lamq =\mu \left(b_0\gbar^2\right)^{-b_1/(2b_0^2)} \rme^{-1/(2b_0\gbar^2)}
 \exp \left\{-\int_0^{\gbar} \rmd x
 \left[\frac{1}{ \betanq(x)}+\frac{1}{b_0x^3}-\frac{b_1}{b_0^2x}
 \right]
 \right\}\,, \label{e_lambdapar}
 \ees
is defined as the integration constant of the solution to the renormalization group equation (RGE)
\bes
\mu {\partial \bar g \over \partial \mu} = \betanq(\bar g)
\label{e_RG}
\ees
for the renormalised coupling $\gbar$ at renormalisation scale $\mu$
with the QCD $\beta$-function
\bes
\betanq(\bar g) & \buildrel {\bar g}\rightarrow0\over\sim &
-{\bar g}^3 \left\{ b_0 + {\bar g}^{2}  b_1 + \ldots \right\}
\enspace ,  \label{e_RGpert} \\ \nonumber
&&b_0=\frac{1}{(4\pi)^2}\left(11-\frac{2}{3}\nf\right)
\enspace ,\quad
b_1=\frac{1}{(4\pi)^4}\left(102-\frac{38}{3}\nf\right) \enspace .
\ees
We shall also make use of the RGI mass
\bes
  M &=& \mbar\,(2 b_0\gbar^2)^{-d_0/(2b_0)}
  \exp \left\{- \int_0^{\gbar} \rmd x \left[{\tau_\tq(x) \over \betanq(x)}
  - {d_0 \over b_0 x} \right] \right\}  \label{e_Mrgi}
\ees
 which appears as an integration constant in the solution of the RGE
\bes
  {\mu \over \mbar} {\partial \mbar \over \partial \mu} &=& \tau_\tq(\bar g) \enspace ,
\ees
\bes
  \label{e_RG_m}
 \tau_\tq(\bar g) & \buildrel {\bar g}\rightarrow0\over\sim &
 -{\bar g}^2 \left\{ d_0 + {\bar g}^{2}  d_1 + \ldots \right\}
 \, , \qquad
 d_0={8}/{(4\pi)^2}\,,
 \label{e_RGpert_m}
 \ees
for the renormalised mass at scale $\mu$.
Amongst different mass definitions, the RGI mass is
distinguished by scale and scheme independence and represents
our choice to discuss mass-dependences. The above holds in any
mass-independent renormalisation scheme.

In the following subsection we discuss how the relation of the $\Lambda$-parameters
of fundamental and effective theory determine the (heavy-) mass dependence
of low energy observables and then turn to the available perturbative information.
This serves to prepare for our subsequent non-perturbative investigation.

\subsection{Non-perturbative matching and mass-dependence}
The leading order (in $1/M$) effective theory describes the fundamental one
at low energy when $\Laml$ has the proper value. In other words, it has
to be chosen as a function of $M$ and $\Lamq$. To make that precise, we specify the $\Lambda$-parameters in units of an arbitrary (but low energy) mass scale $\mscale$.
One may think of a hadron mass or low energy scales such as
$r_0^{-1},\,t_0^{-1/2},\,w_0^{-1}$~\cite{pot:r0,flow:ML,flow:w0}.
The relation between the $\Lambda$-parameters
of fundamental and effective theory may then be written as
\bes
    \label{e:lamrat1}
   \frac{\Lambda_\tl}{\mscale_\tl} =  P^\mscale_{\tl,\tq}(M/\Lamq) \times \frac{\Lambda_\tq}{\mscale_\tq(M)}  \,.
\ees
Since ratios of low energy scales are the same in the leading
order effective theory and in the fundamental theory,\footnote{Such ratios are independent of the value of the coupling constant.}
\bes\label{e:scaleratios}
    \frac{\mscale_\tq(M)}{\mscale_\tq'(M)}
  = \frac{\mscale_\tl}{\mscale_\tl'}  + \rmO((\Lamq/M)^2) \,,
\ees
we may also omit the units and write
\bes
    \label{e:lamrat}
   \Laml =  P_{\tl,\tq}(M/\Lamq) \, \Lamq \,,
\ees
remembering that (non-perturbatively) $P_{\tl,\tq}(M/\Lamq)$
has an  $\rmO((\Lamq/M)^2)$ fuzziness and
that the $\Lambda$'s have to be measured in units of the same
low energy scale in the two theories. One may also read
\eq{e:lamrat} in this way: once the intrinsic
non-perturbative scale of the fundamental theory is specified the equation
determines the one of the effective theory
through the factor $P_{\tl,\tq}(M/\Lamq)$.
Note that by definition the $\Lambda$-parameter of the fundamental theory
does not depend on $M$, but the value of the
dimensionful $\Lambda$-parameter in the effective theory
$\Lambda_\tl$ does depend on it through \eq{e:lamrat}.

Multiplication of \eq{e:lamrat1} with $\mscale_\tq(0) / \Lambda_\tq$ yields the interesting equation
\bes
  {\mscale_\tq(M) \over \mscale_\tq(0)} &=&
    Q^\mscale_{\tl,\tq} \times {P^\mscale_{\tl,\tq}(M/\Lamq)}
    \nonumber \\
    &=& Q^\mscale_{\tl,\tq} \times {P_{\tl,\tq}(M/\Lamq)}
    + \rmO((\Lamq/M)^2) \,.
  \label{e:theequ}
  \ees
with
\bes
   Q^\mscale_{\tl,\tq} =  {\mscale_\tl/ \Laml \over
                   \mscale_\tq(0)/\Lamq }  \,
\ees
defined entirely through the two mass-less theories.
The ratio ${\mscale_\tq(M) \over \mscale_\tq(0)}$
can be computed in the fundamental theory and \eq{e:theequ} is
a consequence of decoupling which can be tested.
We call \eq{e:theequ} \textit{factorisation formula} because it separates the
mass dependence into a ``perturbative'' (see \sect{s:PT}) factor
$P_{\tl,\tq}$ and a non-perturbative factor $Q^\mscale_{\tl,\tq}$ respectively.
In the same loose sense as usually used in factorisation formulae,
the long-distance physics is in $Q$ while the short-distance one is in $P$.
The scale for long/short is given by $1/M$. We have ``perturbative'' in
quotes, because the meaning is not that perturbation theory gives
the complete answer but that it yields an asymptotic expansion.

To simplify the notation, we will from now on omit the subscripts $\tl$, $\tq$ when referring to the quantities
$Q$, $P$.

In phenomenology, \eq{e:theequ} does not seem interesting
since one is usually not interested in the, e.g.,
proton mass at vanishing charm- or bottom-quark mass.
However, in non-perturbative studies of QCD for different flavours the ratio $Q$ is a natural quantity to determine,
and is known to some degree, see below. Testing \eq{e:theequ}
is thus a natural question.
Furthermore, taking
a logarithmic derivative of the nucleon mass w.r.t. the mass $M$ yields the charm content in the nucleon, see Sect.~\ref{s:darkmatter}.

Indeed we will study the mass-scaling function ($P'(x)=\frac{\rmd}{ \rmd x}P(x)$)
\bes\label{e:etaM1}
  \etargi(M) \equiv  {M\over P } \left.{\partial P \over \partial M}\right|_{\Lamq}
  = {M\over \Lamq}{P'\over P}\,,
\ees
which can be computed in perturbation theory when $M$ is
sufficiently large, cf. \cite{Kryjevski:2003mh}.

We can estimate $\etargi$ from the mass dependence of hadronic quantities
by taking the logarithmic derivative in \eq{e:theequ} with respect to the mass
\bes
  \label{e:etaM2}
  {M\over \mscale_\tq } \left.{\partial \mscale_\tq \over \partial M}\right|_{\Lamq} =
    \etargi \,,
    \ees
    where $\mscale_\tq(0)$ and $Q$ drop out. Their uncertainties play no role
and we will therefore be able to make a more stringent comparison between perturbation theory and the full theory.
Of course the $\Lambda^2/M^2$ dependence of $\mscale$ in \eq{e:theequ} is inherited by $\etargi$.

\subsection{Perturbation theory}
\label{s:PT}
We consider a mass-independent renormalization scheme;
whenever we insert perturbative coefficients, it will be in the
$\msbar$-scheme. To simplify notation we use
$\gbar(\mu/\Lambda)\equiv\gbarnq(\mu/\Lamq)$.

\subsubsection{Matching of couplings}

In general form, the relation between the couplings
$\gbar(\mu/\Lambda)$ of the fundamental theory and $\gbareff(\mu/\Lameff)$ of the
leading order effective theory reads
\bes
  \gbareff^2(\mu/\Lameff)=F(\gbar^2(\mu/\Lambda), M/\Lambda)\,.
  \label{e:match1}
\ees
In principle the function $F$ depends on which low energy observable
is matched as discussed in the previous section for $P^\mscale_{\tl,\tq}$. However, that dependence
is only through powers of $\mu_\mathrm{match}/M$, where $\mu_\mathrm{match}$
is the typical energy scale of the matched observable.
In perturbation theory  $(\mu_\mathrm{match}/M)^n$ terms
can uniquely be separated
from the logarithmic $\gbar^2$ terms. Dropping
the power corrections as appropriate for the leading order theory,
the coupling relation (i.e. the function $F$) is thus universal, i.e. independent of the
matching condition.

Choosing the
particular scale $\mu=\mstar$ \cite{Weinberg:1980wa,thresh:BeWe}
in \eq{e:match1},
the first order perturbative correction vanishes
in the $\msbar$ scheme and
we have \cite{Grozin:2011nk,Chetyrkin:2005ia}
\bes
    \label{e:matchg}
   \gbareff^2(\mstar/\Lameff) &=&  \gstar^2\,
   C(\gstar)\,, \quad \gstar \equiv \gbar(\mstar/\Lambda)\,,
   \\ && C(x)= 1+c_2x^4+c_3x^6+c_4x^8+\ldots
   \quad \label{e:matchg_C} \,.
\ees
The scale $\mstar$ is defined such that the running $\msbar$ quark mass fulfills $\mbar(\mstar)=\mstar$.
The two loop coefficient is then given by $c_2 = (\nq-\nl) \, {11\over 72}\,(4\pi^2)^{-2}$.
The coefficients $c_3$ and $c_4$ are known for $\nq-\nl=1,2$ and $\nq-\nl=1$, respectively. They are listed in Appendix \ref{s:coefficients}. One should remember that through
\eq{e_Mrgi}, $\mstar$ and $M$ are in one-to-one relation.

\subsubsection{Mass scaling function $\etargi$}
In order to find the perturbative expansion of $\eta^\mathrm{M}$, \eq{e:etaM1},
we start from the related function (considering
$P(M/\Lambda) = P(M(\mstar,\Lambda)/\Lambda)$)
\bes
  \etam = {\mstar\over P } \left.{\partial P \over \partial \mstar}\right|_{\Lambda}\,,
\ees
which appears upon taking a derivative with respect to the logarithm of
$\mstar$ on
both sides of \eq{e:matchg}. The left hand side yields
\bes
     \mstar {\partial \gbareff^2 \over \partial \mstar }
      =2\gbarnl \betanl(\gbarnl) \,[1 - \etam] \,,
\ees
where we used the matching condition $\gbareff(\mstar/\Lameff)=\gbarnl(\mstar/(P\Lambda))$.
Combined with the straightforward derivative of the right hand side we can solve for $\etam$ and obtain
\bes
    \label{e:etam-matching}
    \etam = 1 - { \betanq(\gstar) \over  \betanl(\gstar\,\Cs(\gstar)) }
      \left[\Cs(\gstar) + \gstar  {\rmd  \over \rmd \gstar}\Cs(\gstar) \right] \,, \quad \Cs(x)=\sqrt{C(x)}\,,
\ees
where we used \eq{e:matchg} to replace $\gbarnl=\gstar\,\Cs(\gstar)$.
Finally, with ${M \over \mstar} {\partial \mstar  \over \partial M} = (1-\tau_\tq(\gstar))^{-1}$,
(see eg. \cite{LH:rainer}, section~3.3.2)
we derive
\bes
  \etaM  = {\etam \over 1 -\tau_\tq(\gstar)}
   \label{e:etaM} \,.
\ees
The first terms in the perturbative expression
\bes
    \label{e:exp-etam}
     \etam = \eta_0 + \eta_1 \gstar^2 + \eta_2 \gstar^4 + \eta_3 \gstar^6 + \eta_4 \gstar^8 + \ldots
\ees
are given by
\bes
     \eta_0 = 1 -{b_0(\nq) \over b_0(\nl)} > 0 \,,\quad
            \eta_1 = (\eta_0-1) \left[\br_1(\nq)-\br_1(\nl)\right] \,,
\ees
with $\br_i(\nf) = b_i(\nf)/b_0(\nf)$. The flavor dependence of the coefficients of the QCD $\beta$-function \eqref{e_RGpert} is made explicit here.
The perturbative expansion of $\etaM$
\bes
\label{e:exp-etaM}
\etaM = \eta_0 + \etaM_1 \gstar^2 + \etaM_2 \gstar^4 + \etaM_3 \gstar^6 + \etaM_4 \gstar^8  + \ldots \,,
\ees
is obtained from \eqref{e:etaM} and the coefficients are given by the recursion
\bes
\etaM_i & = & \eta_i - \sum_{j=0}^{i-1}\,d_j\etaM_{i-1-j}\,.\label{e:etaMrec}
\ees
For example $\etaM_1 = \eta_1 - d_0 \eta_0$, where $d_0$ is the
 universal coefficient of the QCD anomalous dimension \eqref{e_RGpert_m}.
The higher order coefficients $\eta_i$, up to $i=4$, are collected in appendix \ref{s:coefficients}.

We note that for fixed $\nl$ the first two coefficients are exactly proportional to $\nq-\nl$
\bes
	\eta_0 = \frac{2(\nq-\nl)}{33-2\nl} \,,\quad
	\eta_1 = \frac{642(\nq-\nl)}{(4\pi)^2(33-2\nl)^2}\,, \quad
	\etaM_1 = \frac{2(57+16\nl)(\nq-\nl)}{(4\pi)^2(33-2\nl)^2}\,.
\ees
At higher orders this is only true up to small corrections. The dependence on $\nl$ at fixed
$\nq-\nl$ is weak and amounts to a difference of about $20\%$ at leading order between $\nl=0$ and $\nl=3$. In \tab{t:PTcoeffs} we list numerical values for interesting combinations of $\nq$ and $\nl$.

Integrating \eq{e:etaM1} now gives an asymptotic
expression for the mass dependence of non-perturbative
low energy scales $\mscale_\tq$ from perturbation theory ($\logML=\log(M/\Lambda)$)
\bes\label{e:P}
  P = \frac{1}{k} \exp(\eta_0\logML)\; (\logML)^{\etargi_1/(2\bzero)} \times
        \left[1 + \rmO\bigg(\frac{\log(\logML)}{\logML}\bigg) \right],
\ees
where the constant $k$ is fixed by the conventions for the $\Lambda$-parameter
and the RGI mass $M$, which we specified at the beginning of the section, to:
\bes
     \log (k) = {\br_1(\nq) \over 2 \bzero}\log (2) - {\br_1(\nl) \over 2 \bzerop}
                       \log(2\bzero/\bzerop)\,.
\ees
See Appendix \ref{s:asymptotic} for the derivation of \eq{e:P}.
We note that the leading correction
in the expansion \eq{e:P} is $\log(\logML)/\logML$. It contains a term
$\gstar^2\log(\gstar^2)$, cf. \eq{e:lvsgbar}, which makes the convergence
of the expansion slow. Therefore for the numerical evaluation of $P$ we
prefer to use the formula \eq{e:PLambda} which has corrections only in powers
of $\gstar^2$ (no logarithms), see the details in \sect{s:ptaccuracy}.
Accidentally, for the interesting cases, the asymptotic expression
\eq{e:P} for $P$ is dominated by
$\exp(\eta_0\logML) = (M/\Lambda)^{\eta_0}$. This can be seen by the
numerical smallness of $\etaM_1/2\bzero$ and $\log(k)$ in \tab{t:PTcoeffs}.

%%%%%%%%%%%%%%
\begin{table}[t]
 \centering
\begin{tabular}{ccccccc}
\toprule
$\nq$ & $\nl$  & $\eta_0$  & $\eta_1$  & $\etaM_1$ & $\etaM_1/2\bzero$   & $\log(k)$\\
\midrule
2 & 0 & 0.121212 & 0.007467 & 0.001326 & 0.010829 & 0.046655 \\
5 & 3 & 0.148148 & 0.011154 & 0.003648 & 0.037574 & 0.017501 \\
4 & 3 & 0.074074 & 0.005577 & 0.001824 & 0.017284 & 0.012756 \\
5 & 4 & 0.080000 & 0.006505 & 0.002452 & 0.025252 & 0.002622 \\
\bottomrule
\end{tabular}
 \caption{Numerical size of the perturbative coefficients in eqs. \eqref{e:exp-etam}, \eqref{e:exp-etaM} and \eqref{e:P}.
}
 \label{t:PTcoeffs}
\end{table}
%%%%%%%%%%%%%%

\subsection{Accuracy of perturbation theory \label{s:ptaccuracy}}

\begin{figure}[ht]
\includegraphics[width=0.49\textwidth]{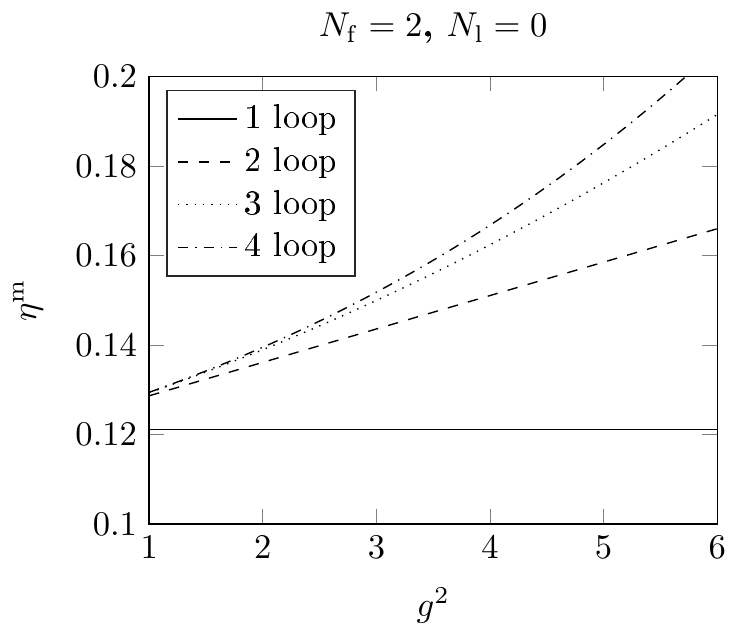}
\includegraphics[width=0.49\textwidth]{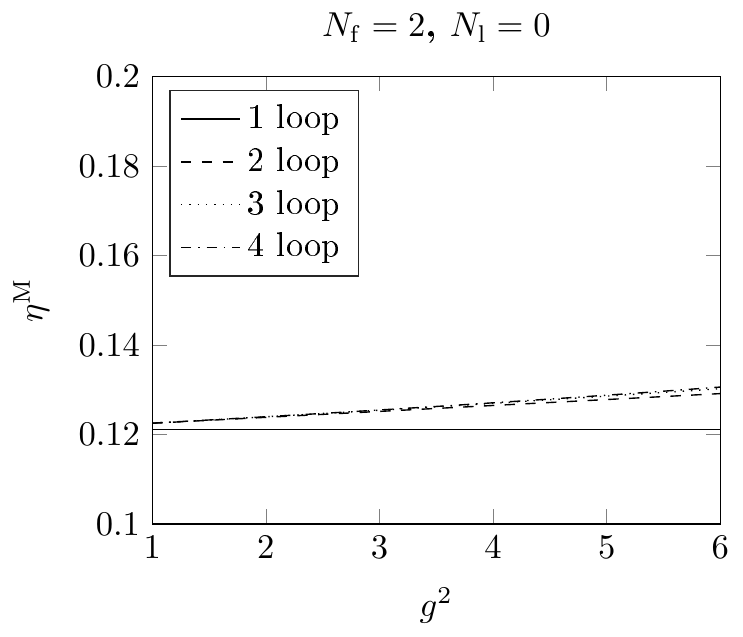}
\caption{The functions $\etam(g^2),\etargi(g^2)$ for the case $\nq=2,\nl=0$.
The number of loops corresponds to the highest loop order of the $\beta$ function which is used.}
\label{f:eta2}
\end{figure}

\begin{figure}[ht]
\includegraphics[width=0.49\textwidth]{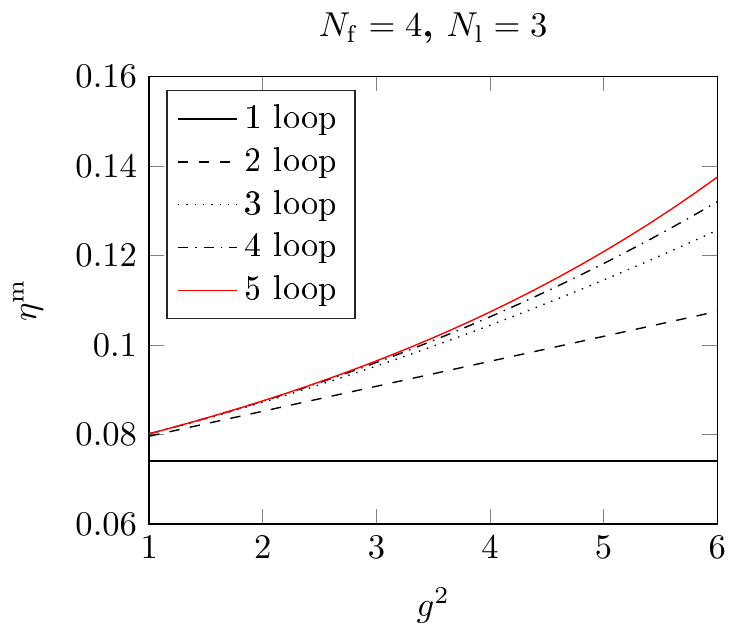}
\includegraphics[width=0.49\textwidth]{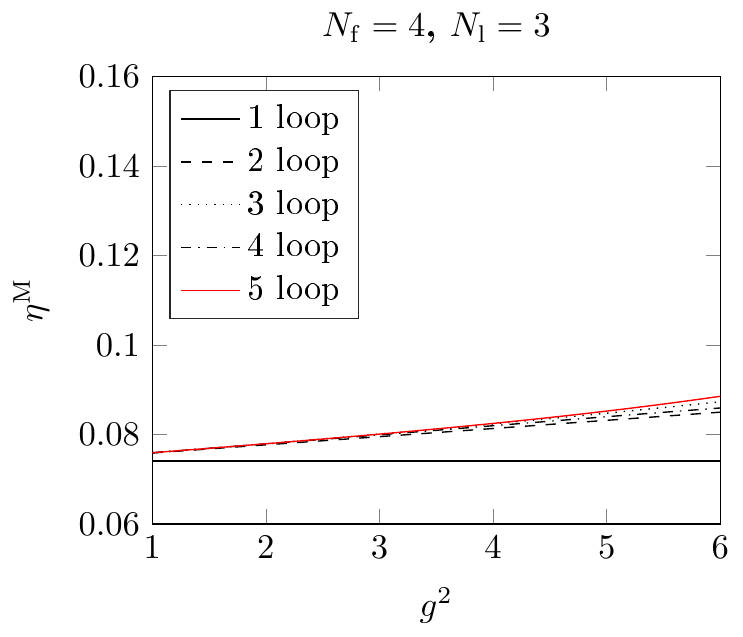}
\caption{The functions $\etam(g^2),\etargi(g^2)$ for the case $\nq=4,\nl=3$.
The number of loops corresponds to the highest loop order of the $\beta$ function which is used.}
\label{f:eta4}
\end{figure}

\begin{figure}[ht]
\includegraphics[width=0.49\textwidth]{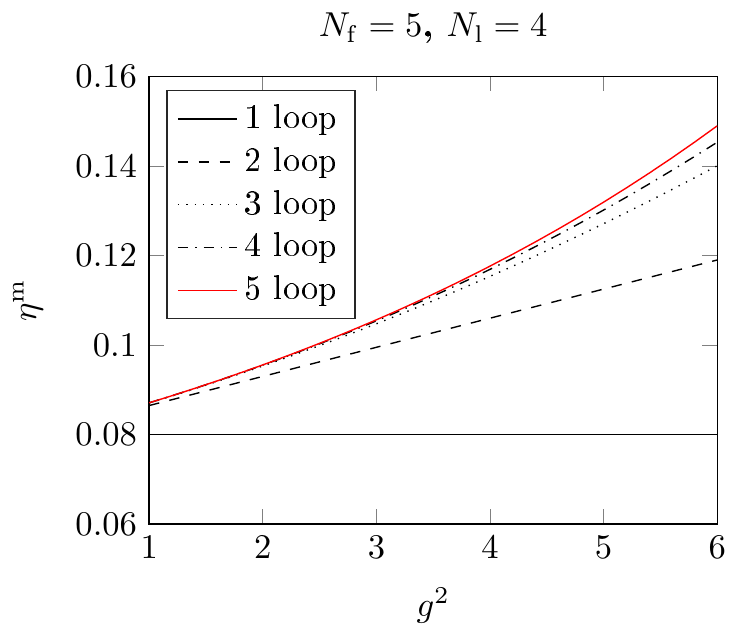}
\includegraphics[width=0.49\textwidth]{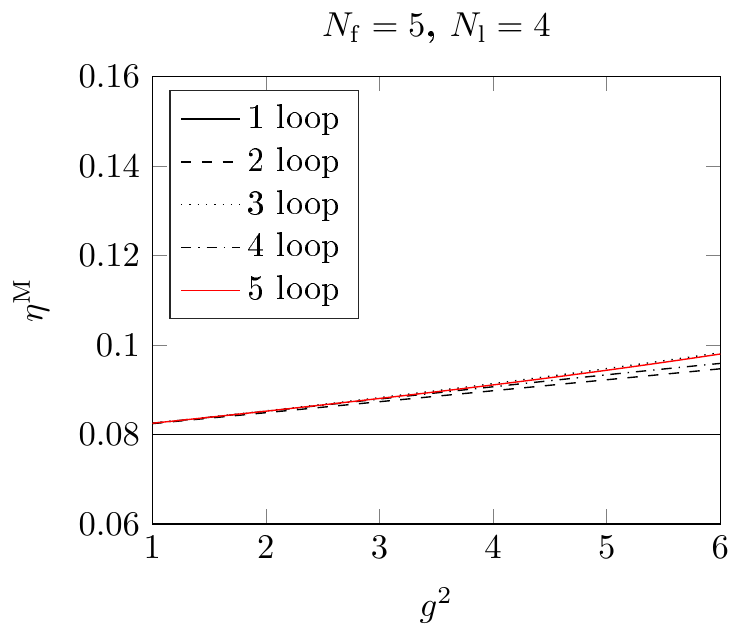}
\caption{The functions $\etam(g^2),\etargi(g^2)$ for the case $\nq=5,\nl=4$.
The number of loops corresponds to the highest loop order of the $\beta$ function which is used.}
\label{f:eta5}
\end{figure}

A consistency check on the applicability of
perturbation theory is the comparison of
different orders. Indeed, figures
\ref{f:eta2}-\ref{f:eta5} show that
higher orders
do not contribute very much, in particular
when one uses the mass dependence in terms of the RGI mass, $\etargi$.
This also suggests that it is an advantage to consider
the perturbative prediction for $P$ in terms of $M/\Lambda$ instead
of working with $\mstar/\Lambda$. We have worked with $M/\Lambda$
in \cite{Bruno:2014ufa} and will do so below in our comparison to
a non-perturbative investigation.

Details for $\etam$ and $\etargi$ are seen in figure
\ref{f:eta2}-\ref{f:eta5}. In the legends of the plots the number of loops
corresponds to the highest loop order of the $\beta$ function which is used.
We note
that in the right plot of figure \ref{f:eta4} the 5-loop
correction is larger in magnitude than the 4-loop correction for $g^2\gtrsim 3$.
But the corrections are amazingly small.

\begin{figure}[ht]
\centering
\includegraphics[width=0.49\textwidth]{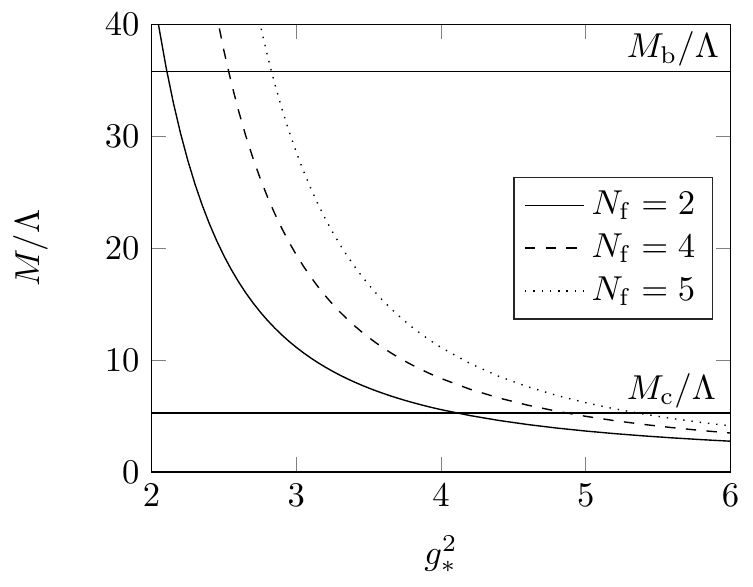}
\includegraphics[width=0.49\textwidth]{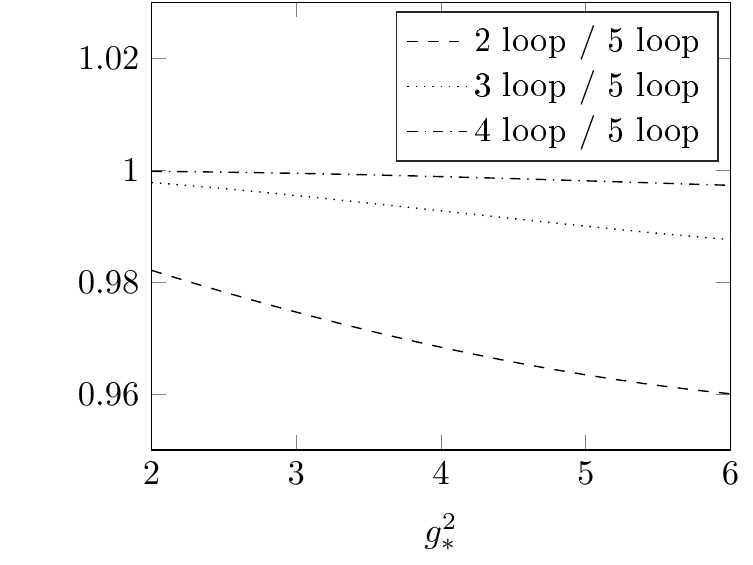}
\caption{Left: The relation between $M/\Lambda$ and $\gstar$ at 5-loop.
Right: The 2, 3 and 4-loop relation divided by the 5-loop one for the case of $\nq=2,\nl=0$.}
\label{f:ML}
\end{figure}

Renormalization group improved perturbative predictions for the
function $P(M/\Lambda)=\Laml/\Lamq$ can be obtained from (cf.~\eq{e_lambdapar})
\bes
\label{e:PLambda}
  P(M/\Lambda) = \exp\left\{ I_g^{\tl}(\gstar\,\Cs(\gstar)) -  I_g^{\tq}(\gstar)\right\} \,,
\ees
where
\bes
 \exp( I_g^i(\gbar)) &=&\left(b_0(N_i)\gbar^2\right)^{-b_1(N_i)/(2b_0(N_i)^2)} \rme^{-1/(2b_0(N_i)\gbar^2)}\\
          && \times
           \exp \left\{-\int_0^{\gbar} \rmd x
          \left[\frac{1}{ \beta_i(x)}+\frac{1}{b_0(N_i)x^3}-\frac{b_1(N_i)}{b_0(N_i)^2x}
          \right]
          \right\} \enspace . \label{e:lambdapar}
\ees
The coupling $\gstar=\gbar(\mstar)$ is obtained from inverting
\bes
%{\Lambda \over M} &=&  \exp \left\{-\int^{\gstar(M/\Lambda)} \rmd x\;
%\frac{1-\tau_\tq(x)}{ \beta_\tq(x)}          \right\} \,,
    {\Lambda \over M} =
    \frac{\left(b_0\gbar^2\right)^{-b_1/(2b_0^2)}}{(2 b_0\gbar^2)^{-d_0/(2b_0)}}
    \rme^{-1/(2b_0\gbar^2)}
    \exp \left\{-\int_0^{\gstar(M/\Lambda)} \rmd x\;
    \left[\frac{1-\tau_\tq(x)}{ \beta_\tq(x)}
      +{1 \over b_0 x^3} - {b_1 \over b_0^2 x} + {d_0 \over b_0 x}\right]
      \right\} \,, \nonumber\\
  \label{e:gstarM}
\ees
where $M$ is the RGI mass corresponding to $\mstar$. For this equation we have
combined eqs. \eqref{e_lambdapar} and \eqref{e_Mrgi} using $\mu=\mbar=\mstar$. For
reference the resulting relation is plotted in the left panel of \fig{f:ML}
together with the values for $\Mc/\Lambda$ and
$\Mb/\Lambda$ which were obtained from the PDG values \cite{Olive:2016xmw} for $\mbar_\mathrm{c}/\Lambda$ and $\mbar_\mathrm{b}/\Lambda$, and inverting \eq{e_lambdapar}.
Of course, in case of the charm quark $\nq=4$ and in the case of the bottom quark $\nq=5$ were
used.

\begin{figure}[ht]
\includegraphics[width=0.49\textwidth]{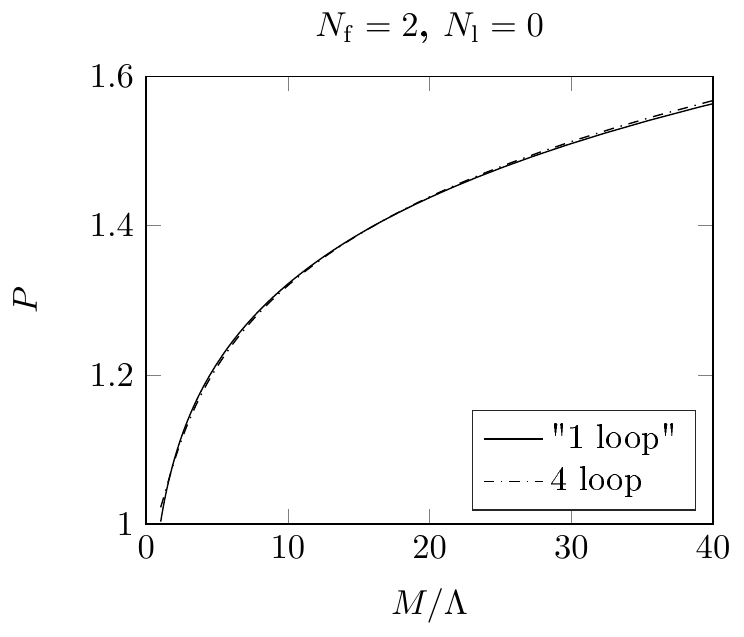}
\includegraphics[width=0.49\textwidth]{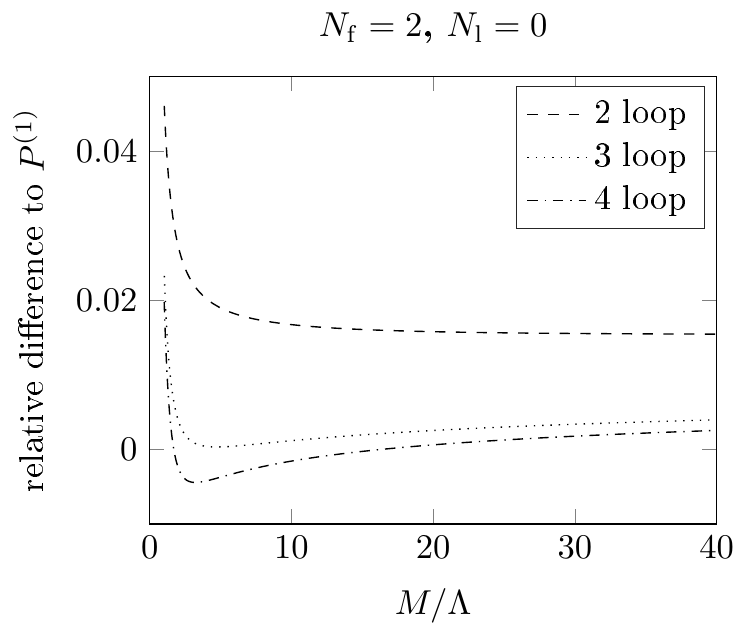}
\caption{The mass-dependence $P$ at 1-loop formula and at 4-loop (left) as
well as 2,3,4-loop correction normalised to the 1-loop approximation
(right) for the  case $\nq=2,\nl=0$.}
\label{f:LL2}
\end{figure}

\begin{figure}[ht]
\includegraphics[width=0.49\textwidth]{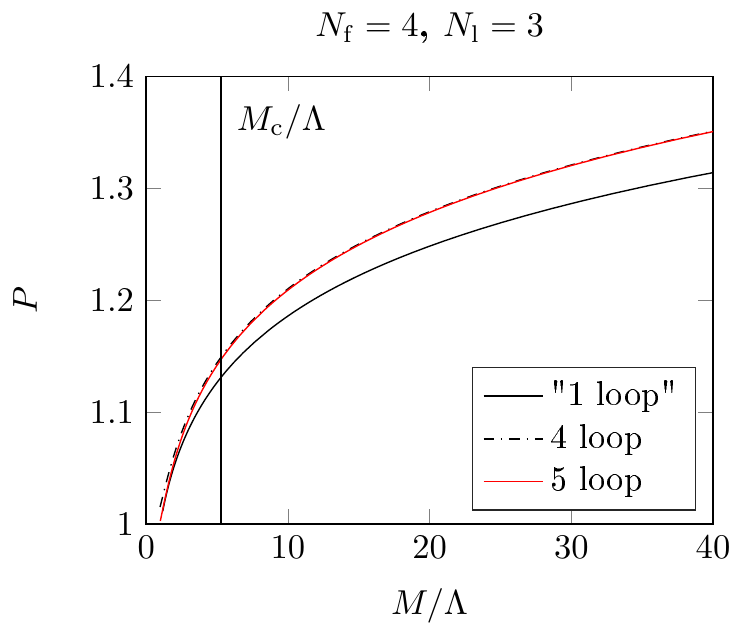}
\includegraphics[width=0.49\textwidth]{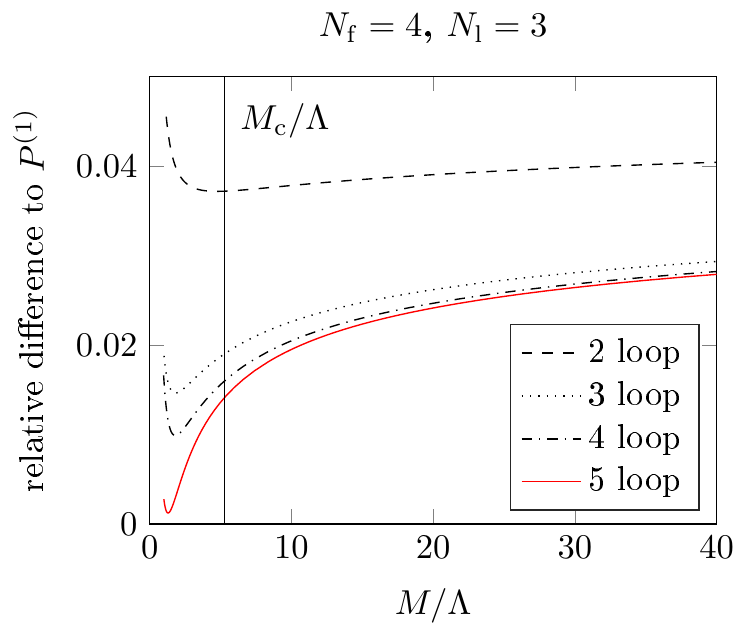}
\caption{The mass-dependence $P$ at 1-loop formula and at 4,5-loop (left) as
well as 2,3,4,5-loop correction normalised to the 1-loop approximation
(right) for the  case $\nq=4,\nl=3$.}
\label{f:LL4}
\end{figure}

\begin{figure}[ht]
\includegraphics[width=0.49\textwidth]{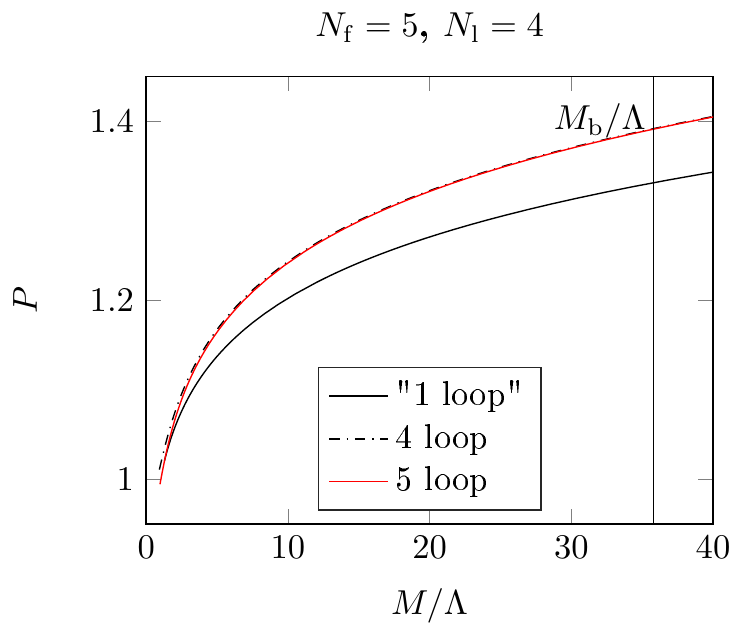}
\includegraphics[width=0.49\textwidth]{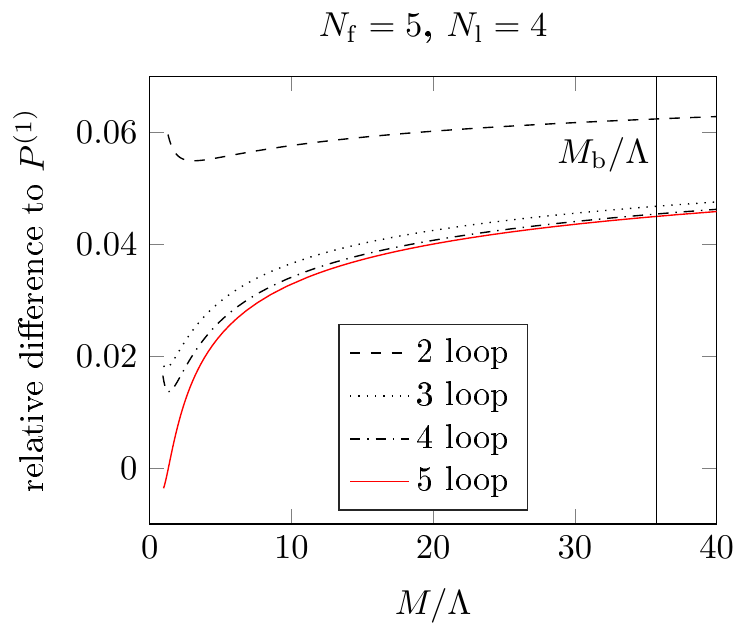}
\caption{The mass-dependence $P$ at 1-loop formula and at 4,5-loop (left) as
well as 2,3,4,5-loop correction normalised to the 1-loop approximation
(right) for the  case $\nq=5,\nl=4$.}
\label{f:LL5}
\end{figure}

The predictions for different orders of perturbation theory
are very close to the unsystematic
one-loop ``approximation'', $P^{(1)} = (M/\Lambda)^{\eta_0}$,
as long as $M/\Lambda < 30$ or so and the number of flavors is small.
This is accidental. In figure \ref{f:LL2}--\ref{f:LL5} we plot the one-loop ``approximation''
and the 4-loop result on the left and the relative correction
\bes
   (P - P^{(1)}) / P^{(1)}\,.
\ees
at 2,3,4-loop on the right. When it is available we also add the 5-loop result.
In this comparison, when we consider at least
2-loop precision, we always work to a consistent order
in the renormalization group functions. Note that we truncate the
renormalization group functions $\beta,\tau$ in the integrals \eq{e:lambdapar},
\eq{e:gstarM} and 2-loop accuracy means, e.g.,
\bes
   \frac{1-\tau(x)}{ \beta(x)} = -{1 \over x^3} \left[ {1 \over b_0 + b_1 x^2} + x^2 {d_0 \over b_0} \right]\,.
\ees
 The function $C(g)$ only enters at
 3-loop precision since $c_1=0$.
 It is only needed for the upper integration limit in \eq{e:PLambda}
 and there we compute explicitely $\Cs(\gstar)=\sqrt{C(\gstar)}$.

In the numerical results we observe in particular that
for the phenomenologically relevant case of $\nq=5,\nl=4$, the
3-loop contribution (difference 3-loop to 2-loop) is around 2\% while the 4- and 5-loop ones are
then nice and small, see the right plot in \fig{f:LL5}. Judging by perturbation theory alone,
the perturbative predicition for decoupling the b-quark
should be very reliable. Also for the other phenomenologically relevant case
of decoupling the c-quark ($\nq=4,\nl=3$) perturbation theory appears to work
quite well.
%%%%%%%%%%%%%%
\begin{table}[t]
 \centering
\begin{tabular}{ccccccc}
\toprule
$\nq$ & $\nl$ & 1 loop & 2 loop & 3 loop & 4 loop & 5 loop \\
\midrule
2 & 0 & 1.2319 & 1.2546 & 1.2170 & 1.2084 & - \\
4 & 3 & 1.1448 & 1.1875 & 1.1552 & 1.1492 & 1.1468 \\
5 & 4 & 1.3413 & 1.4255 & 1.3947 & 1.3918 & 1.3913 \\
\bottomrule
\end{tabular}
 \caption{Perturbative values of $P_{\tl,\tq}$ defined in \eq{e:lamrat} for
various cases of interest, see main text for details.
}
 \label{t:P}
\end{table}
%%%%%%%%%%%%%%

These curves suggest that perturbative decoupling introduces only
errors at the sub-percent level for the ratios of Lambda parameters,
once perturbation theory applies at all.
In \tab{t:P} we list the values of $P$ computed from \eq{e:PLambda} using
different orders of perturbation theory. We evaluate $P$ at an argument
$M/\Lambda$ which depends on $\nq$ and $\nl$.
For $\nq=2,\nl=0$ we obtain $M/\Lambda$ from the PDG value for
$\mbar_\mathrm{c}$ \cite{Olive:2016xmw} and $\Lambda_2=310\,\mathrm{MeV}$
from~\cite{alpha:lambdanf2}.
In this case there is no 5-loop result because the coefficient $c_4$ is
not known.
For $\nq=4,\nl=3$ and $\nq=5,\nl=4$ we use the PDG values for $\Mc/\Lambda$ and
$\Mb/\Lambda$ as explained above.

\section{Non-perturbative investigation
for $\nq=2 \,\to\, \nl=0$}
\label{s:simulation}

We investigate a model,
namely QCD with $\nq=2$ heavy, mass-degenerate quarks. The decoupling is
then $2\to0$ and the Lagrangian of the effective theory,
$\lag{\rm dec}$, is the Yang-Mills one
up to $1/M^2$ corrections. 
We target the RGI quark mass values (see below)
\bes
 \label{e:targetM}
\frac{M_\mathrm{targ}}{\Lambda} = 0.59\,,\; 1.28\,,\; 2.50\,,\; 4.87 \,,\; 5.7781\,.
\ees
Using $\Lambda\equiv\Lambda_2=310\,\MeV$ from \cite{alpha:lambdanf2} their
physical values are approximately $M_\mathrm{targ}=$0.2, 0.4, 0.8, 1.5, 1.8$~\GeV$. 
The value $M_\mathrm{targ}/\Lambda=4.87$ corresponds to
the RGI charm quark mass $\Mc$ from \cite{Heitger:2013oaa}
in agreement with \cite{Olive:2016xmw}
within the present uncertainties.
However, for our model study the exact value is not
important.

\subsection{Low energy observables}
\label{s:obs}

In principle any low-energy hadronic scale $\mscale(M)$ can be used to study decoupling, but
in practice some choices are far superior to others. Ideally we look for a quantity that is
easily non-perturbatively renormalizable, well defined in both full and effective
theory, has controllable lattice artifacts, is cheap to compute and can be determined with a
high precision. Since in our case the effective theory has no fermionic content, we are restricted to
purely gluonic observables. Glueball masses would be natural candidates. However, it is difficult to 
determine them precisely enough. 
Hadronic scales derived from the static quark potential fulfill all criteria and have been popular for many years. If $F(r)$ denotes the force between two static quarks (defined in terms of the fundamental
Wilson loop), a
distance $r_x$ can be defined implicitly~\cite{pot:r0} by choosing a number $c$ and solving
\bes
   \label{e:r0r1}
   r_x^2 F(r_x) = c\, .
\ees
The choices $r_0 \Leftrightarrow c=1.65$~\cite{pot:r0} and $r_1 \Leftrightarrow c=1.0$~\cite{pot:r1} have become standards. In a lattice calculation the latter has a better statistical 
precision, but larger lattice artifacts. Moreover we expect decoupling to be more precise for the longer distance, $r_0$.

In recent years, these scales have been largely replaced by scales based on the gradient flow \cite{flow:ML,flow:Herbert}.
The gauge field $A_\mu$ is used as an initial condition in a flow equation, that describes the
relaxation of 
a field $B_\mu$ as a function of a flow time $t$.
\bes
   \label{e:flow}
   \partial_t B_\mu = D_\nu G_{\nu\mu}\, , \qquad B_\mu\bigr|_{t=0} = A_\mu\, .
\ees
The field strength tensor $G_{\nu\mu}$ and the covariant derivative $D_\nu$ are defined in the 
usual way, but at flow time $t$. The crucial observation, that correlators of the $B_\mu$ fields
at finite flow time are renormalized quantities~{\cite{flow:LW}}, 
allowed to introduce a family of scales. The 
definition of scales $\sqrt{t_0}$~\cite{flow:ML}, $\sqrt{t_c}$ and $w_0$~\cite{flow:w0} 
is based on the dimensionless combination
\bes
   \label{e:t2E}
   \flowE(t) = t^2 \left\langle \frac{1}{4} G_{\mu\nu}^aG_{\mu\nu}^a \right\rangle\, ,
\ees
together with
\begin{eqnarray}
   \flowE(t_0)         &=& 0.3 \label{e:t0def}\, , \\
   \flowE(t_c)         &=& 0.2 \label{e:tcdef}\, , \\
   w_0^2\flowE'(w_0^2) &=& 0.3 \label{e:w0def}\, .
\end{eqnarray}
In our simulations we compute the hadronic scales
\bes
 \label{e:hadscales}
\mscale(M) = \frac{1}{r_0}\,,\; \frac{1}{\sqrt{t_0}} \,,\; \frac{1}{\sqrt{t_c}} 
\,,\; \frac{1}{w_0} \,.
\ees 

The rest of this section contains technical details about the lattice
  simulations. It can be omitted if one is only interested in the
  physical results presented in \sect{s:np}.

\subsection{Fixing the RGI parameters of the theory
and details of the simulations}

\subsubsection{Discretization}

We use Wilson's plaquette gauge
action \cite{Wilson} and include quarks treated
with two discretizations:
O($a$) improved Wilson fermions~\cite{Sheikholeslami:1985ij,Luscher:1996sc} and twisted mass \cite{tmqcd:pap1}
Wilson fermions at maximal twist.
For both actions
the clover term~\cite{Sheikholeslami:1985ij,Luscher:1996sc} has the
non-perturbatively determined improvement coefficient $c_{\rm sw}$~\cite{impr:csw_nf2}.
Twisted mass fermions at maximal twist are automatically O($a$) improved
\cite{tmqcd:FR1} also without a clover term. However, 
with the clover term added our two discretizations
have a common chiral limit in a finite volume (see $L_1$ below).
Furthermore the clover term reduces 
O($a^2$) lattice artifacts as it was shown for example in \cite{Dimopoulos:2009es}.

In appendix \ref{s:tables} we list the ensembles generated with standard Wilson
fermions in \tab{t:ens-Wilson} and with twisted mass Wilson fermions in
\tab{t:ens-tm}.
The twisted mass simulations are the same as in \cite{Knechtli:2017xgy}.

We determine the lattice spacings through the scale $L_1$
\cite{hqet:paramnf2,alpha:lambdanf2}, which is
defined by $\gbarSF^2(L_1)=4.484$ through the
so-called Schr\"odinger Functional coupling
at zero quark mass and in a finite volume of
size $L_1^4$. Note that in this situation 
the two discretizations are identical. Thus at a given
gauge coupling  $\beta=6/g_0^2$ they have 
one and the same lattice spacing. 
The values of $L_1/a$ and the corresponding
lattice spacings are listed in \tab{t:scale-L1}.

\subsubsection{$\Oa$ improvement and finite size effects}
\label{s:improvement}

$\Oa$ improvement of quark mass effects requires to keep the improved bare coupling
$\tilde{g}_0^2=(1+\bg(\nq)\,{a\mq})\,g_0^2$ fixed, where 
$\mq=1/(2\kappa) - 1/(2\kappa_\mathrm{c})$ is the bare subtracted standard mass.
Twisted mass fermions at maximal twist have $\mq=0$ and therefore the
improved coupling is $\tilde{g}_0=g_0$.
Instead our simulations with standard Wilson fermions were done at fixed 
$g_0$ (and not $\tilde{g}_0$).
We correct for the resulting $\rmO(am)$ effects in the lattice spacing by 
decreasing the values of $a\mscale(M)$ using the 1-loop result 
$\bg(\nq)=0.01200\, \nq\, g_0^2 $ \cite{pert:1loop,impr:pap1} and the 1-loop $\beta$-function.
For $\mscale=1/\sqrt{t_0(M)}$ these effects shift the value of $\sqrt{t_0(M)}/a$  
according to
\begin{equation}\label{e:bgcorr}
   \left. {\sqrt{t_0(M)} \over a}\right|_{\tilde g_0} \approx 
  \left.{\sqrt{t_0(M)} \over a}\right|_{g_0}\,\times\,\left[1+{0.01200\, \nq \over 2b_0(\nq)}\;a\mq\right]\,.
\end{equation}
We use $a\mq = am/(Zr_{\mathrm{m}})$ and the factor $Zr_{\mathrm{m}}$ is
taken from \cite{alpha:lambdanf2} (at $6/g_0^2=5.7$ we get
$Zr_{\mathrm{m}}=1.194$ from a Pad\'e fit). Here $am$ denotes the PCAC mass. 
We added in quadrature 100\% of the correction to the errors as an estimate of
unknown $\rmO(g_0^4)$ terms in $\bg$.
After the corrections the values of $a\mscale(M)$ correspond to simulations
performed at $\beta=6/\tilde{g}_0^2$.

Our volumes are such that the lightest pseudo-scalar mass times the box size is
$\mps L \ge 7.4$ and $L/\sqrt{t_0(M)} \ge 12$ and $L/r_0(M) \ge 3.8$.
At our largest masses the situation is comparable to the pure gauge theory,
where significant finite volume effects can be excluded for a lattice size
$L\approx 4r_0=2.0\fm$. Approximate decoupling of the 
heavy quarks means that also our finite mass simulations
are practically free of finite volume effects.

\subsubsection{Quark masses}
\label{s:quark_masses}

{Before taking the continuum limit, we non-perturbatively 
fix the value of the RGI quark mass $M$ in units of the
$\Lambda$ parameter through the following steps.
We take the $\Lambda$ parameter to be defined in the 
$\msbar$ scheme while the RGI mass $M$ is independent of the scheme.

In the case of standard Wilson fermions
the renormalized quark mass in lattice units
$a\mbarsf(L_1)$ at length scale $L_1$ is defined 
by $a\mbarsf(L_1)=\za/\zp(L_1)\, am$, where the renormalisation
factor $\zp(L_1)$ is defined in the Schr\"odinger Functional scheme
as in \cite{alpha:lambdanf2} and also the 
details of the definition of $m$ are found there. The axial current renormalization factor, $\za$,
is fixed by a chiral Ward identity \cite{DellaMorte:2008xb} \footnote{
A more precise determination of $\za$ became recently available \cite{DallaBrida:2018tpn}.}. 
For the determination of the PCAC mass $am$
we use our publicly available program\footnote{
It is available at {\tt https://github.com/to-ko/mesons}.}.
The ratio $M/\Lambda$ is then obtained from
\begin{equation}\label{e:MoL}
  \frac{M}{\Lambda} = a\mbarsf(L_1) \times M/\mbarsf(L_1) \times
  \frac{(L_1/a)}{(\Lambda L_1)}\,,
\end{equation}
where we take
$M/\mbarsf(L_1)=1.308(16)$ from \cite{mbar:nf2,alpha:lambdanf2} and
$\Lambda\,L_1=0.649(45)$ from \cite{alpha:nf2}.
The values of the PCAC mass $m$ and of $M/\Lambda$ are tabulated in \tab{t:ens-Wilson}.
The accuracy of $M/\Lambda$ is around 7\% with an error dominated
by the one of $\Lambda\,L_1$. Thus, ratios of masses $M_1/M_2$
or equivalently logarithmic derivatives with respect to masses 
are known significantly more precisely.

In the case of twisted mass fermions at maximal twist the
difference is that
the renormalized quark mass $a\mbarsf(L_1)$ is calculated through 
$a\mbarsf(L_1)=a\mu/\zp(L_1)$, where $a\mu$ is the twisted mass parameter.
The ratio $M/\Lambda$ is again
obtained from \eq{e:MoL}. For twisted mass fermions we actually invert
\eq{e:MoL} to determine the twisted mass parameter corresponding to
given values of $M/\Lambda$ which are tabulated in \tab{t:ens-tm}.

\subsubsection{Hadronic scales on the lattice}
\label{s:hadscales}

In our simulations we measure the observables discussed in
\sect{s:obs}. Various details concerning their computation in the 
discretized theory are as follows.

The clover (symmetric) definition of the action density $E$ is used
in \eq{e:t2E} and we use the Wilson-flow equation, cf. \cite{flow:ML}.

The scale $r_0$ is defined with the ``HYP2'' action 
for the static quarks 
\cite{DellaMorte:2005yc}. It is
determined with our publicly available program\footnote{
It is available at {\tt https://github.com/bjoern-leder/wloop}.} 
following the details explained in Ref.~\cite{pot:nf2}.
We use a variational basis with
up to four levels of spatial HYP smearing~\cite{HYP}
to construct a matrix of Wilson loops.
Due to the open boundary conditions, Wilson loops are averaged only
in a temporal region sufficiently far away from the boundaries to exclude
contaminations from boundary effects.
The static potential as a function of $r$ is obtained
by solving the generalised eigenvalue problem as discussed in Ref.~\cite{pot:nf2}.

Hadronic scales such as $t_0$ are non-linear functions of one or more Monte-Carlo averages of
``primary observables'' $\langle \mathcal{O}_1\rangle,\ldots,\langle \mathcal{O}_{N_{\rm ob}}\rangle$,
like for instance 
the action densities at different flow times. The derivative of such a function with respect 
to the twisted mass, as needed for the MC evaluation 
of $\etargi$ (below in~\eq{e:etaMderivt0}), is in general given by
\begin{equation}
   \frac{\mathrm{d} f(\langle \mathcal{O}_1\rangle,\ldots,\langle\mathcal{O}_{N_{\rm ob}}\rangle,\mu)}{\mathrm{d} \mu} = 
   \sum\limits_{i=1}^{N_{\rm ob}} \frac{\partial f}{\partial \langle \mathcal{O}_i\rangle}\, 
                                  \frac{\mathrm{d} \langle \mathcal{O}_i\rangle }{\mathrm{d}\mu} 
   + \frac{\partial f}{\partial \mu}\, 
\end{equation}
and the derivative of a primary observable $\mathcal{O}$, 
\begin{equation}
\frac{\mathrm{d}\ev{\mathcal{O}}}{\mathrm{d}\mu} =
-\ev{\frac{\mathrm{d}S}{\mathrm{d}\mu}\,\mathcal{O}}
+\ev{\frac{\mathrm{d}S}{\mathrm{d}\mu}}\,\ev{\mathcal{O}}
+\ev{\frac{\mathrm{d}\mathcal{O}}{\mathrm{d}\mu}} \,.
\end{equation}
For most observables $\frac{\partial f}{\partial \mu}$ and $\frac{\mathrm{d}\mathcal{O}}{\mathrm{d}\mu}$ are absent. 
The derivative of the action is given by $\mathrm{d}S/\mathrm{d}\mu = i a^4\sum_x \bar \psi(x)\gamma_5 \tau^3 \psi(x)$.
In cases like ours, where the observables do not contain fermionic fields, no new Wick contractions arise
in the first term, and one simply needs to determine the observable and the action-derivative on each configuration 
and compute their connected correlation. For the action-derivative we write (cf.~\cite{Jansen:2008wv})
\begin{eqnarray}
\left\langle\frac{\mathrm{d}S}{\mathrm{d}\mu}\right\rangle &=& i a^4\sum_x\left \langle \mathrm{tr}
   \left[(D_d(x,x)^{-1}-D_u^{-1}(x,x))\gamma_5 \right]\right\rangle^{\mathrm{gauge}} \nonumber \\ 
   &=& -2\mu a^8\sum_{x,y}\ev{\tr\left[{D_u^{-1}}^\dagger(x,y)D_u^{-1}(x,y)\right]}^{\mathrm{gauge}}\, ,
\end{eqnarray}
where in the last step a property of the twisted mass Dirac operators $D_{u,d}$ (for up and down quark),
$D_u-D_d = 2i\gamma_5 \mu$, was exploited, leading to an expression that has a smaller variance, when
the trace is estimated stochastically. A stochastic estimation is necessary to avoid a 
full matrix inversion, and amounts to solving equations $D_u \xi = \eta$, with 4D noise spinors
$\eta$, for $\xi$ and a subsequent dot product $\xi\cdot\xi$. We find that different noise
distributions (e.g. normal or $U(1)$-noise) yield a similar variance, and further refinements like
spin or color dilution~\cite{Bernardson:1993he} do not pay off. Not many noise-sources are needed for the final error to be close to
the limiting error due to gauge field fluctuations. In our measurements we settle for 64 $U(1)$ noise 
spinors per configuration.

\subsubsection{Simulation algorithms}

In the case of standard Wilson fermions,
part of the simulations are performed using periodic boundary conditions 
(except for anti-periodic boundary conditions in temporal direction for 
the fermions) and the MP-HMC algorithm \cite{lat10:marina}.
In order to avoid the freezing of the topological charge (see also next section),
for simulations with $t_0/a^2>5.5$ \cite{algo:csd,Bruno:2014ova} we adopt 
open boundary conditions in time and
use the publicly available openQCD package\footnote{
  http://luscher.web.cern.ch/luscher/openQCD/} \cite{algo:openQCD}.
We set the boundary improvement coefficients to their 
tree-level values $c_\mathrm{G}=1$ and $c_\mathrm{F}=1$.
In both cases the fermion determinant is Hasenbusch-factorized \cite{Hasenbusch:2001ne} using a splitting
in two factors, thus two pseudo-fermion fields are needed and a hierarchical numerical integrator is employed 
(Leapfrog and Omelyan-Mryglod-Folk integrator schemes are used at the different levels). The trajectory length is
always set to $2.0$ and configurations and measurements are separated by at least
four trajectories. Most computer resources are spent in the solution of the Dirac
equation with the smallest mass. For $M/\Lambda>1$ we use the SAP preconditioned GCR algorithm \cite{algo:L1a} while
for $M/\Lambda<1$ it is profitable to use a multigrid solver \cite{Frommer:2013fsa}, which is implemented
as the two-grid ``locally deflated'' solver in the openQCD package since version 1.2.
The cost of the simulations is low compared to
simulations in the chiral regime.

In the case of twisted mass fermions we use a version of openQCD, in which
the SAP preconditioner can have a different value of $\mu$ than the
simulated one. In the preconditioner the twisted mass term is defined only
on the even sites.
We achieve
a significant speed up of the SAP preconditioned GCR algorithm by choosing
a value of $\mu$ for the SAP preconditioner which is larger by approximately
a factor 6 than the simulated one (the multi-grid inverter of
\cite{Alexandrou:2016izb} implements a similar strategy inspired by our findings).

Open boundary conditions are used as specified above.
In this setup the Wilson--Dirac operator has two mass parameters, the standard
bare quark mass $m_0$ and 
the twisted mass $\mu$. Maximal twist means that $m_0$ is
set to its critical value $m_c$ which corresponds to
the vanishing of the current (PCAC) quark mass. We extracted 
the critical mass from table 13 in \cite{alpha:lambdanf2},
interpolating the data to the desired $\beta$ values by
a Pad{\'e} fit in $g_0^2=6/\beta$ of the form
\begin{equation}
 am_c(g_0) = u_1\,g_0^2+g_0^4\,\frac{\sum\limits_{k=0}^{3}u_{2+k}\,g_0^{2k}}{1+u_d\,g_0^2}
\end{equation}
where the coefficients $u_1$ and $u_2$
coincide with two-loop perturbation theory \cite{mcrit:2loop}.
The values of the hopping parameter $\kappa=1/(2am_c+8)$ are listed in
\tab{t:ens-tm}.

\subsubsection{Autocorrelation times and error analysis}
\begin{figure}[t]\centering
  \includegraphics[width=.8\textwidth]{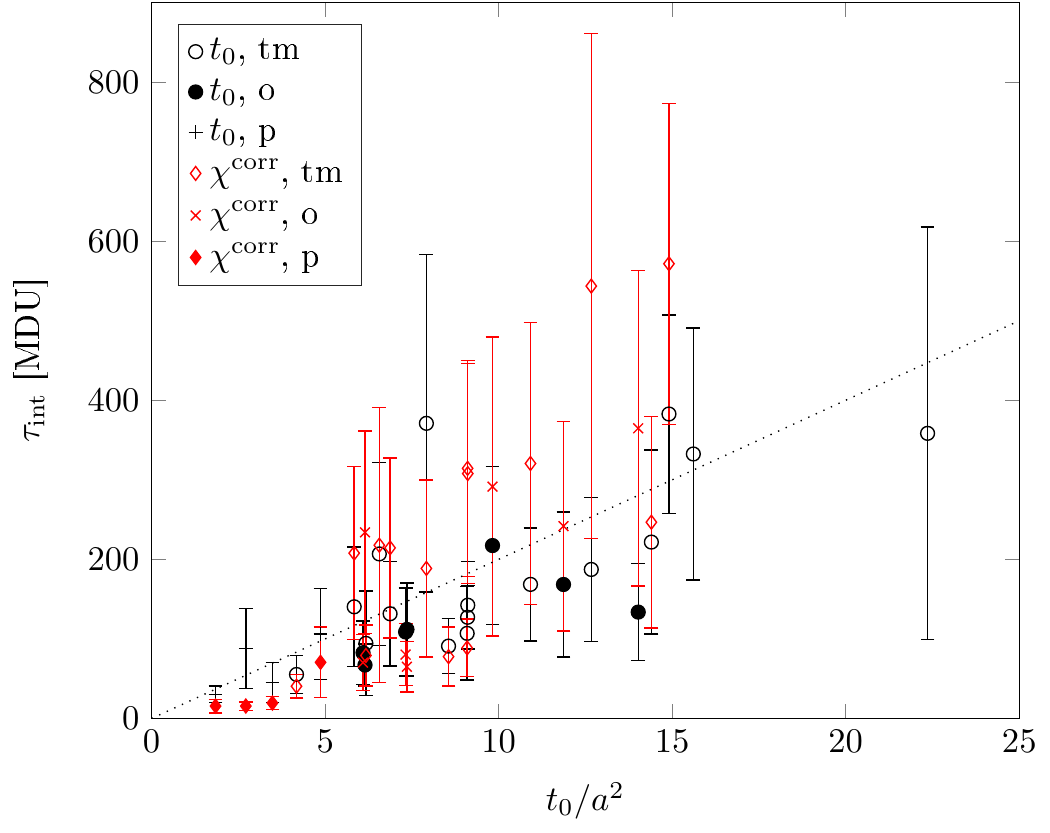}
  \caption{Autocorrelation times derived from observables which are expected to have large
           overlap with the slowest modes in the simulation are plotted as a function of
           $t_0(M)/a^2$. The dotted line represents \eq{e:tauexp}.
          }
  \label{f:taui}
\end{figure}
We measure the integrated autocorrelation time $\tau_{\rm int}$ for all measured quantities including the
hadronic scales, the PCAC mass and additionally the topological susceptibility.
We find the largest $\tau_{\rm int}$ for the scale $t_0$ and for
the topological susceptibility $\chi^\mathrm{corr}$  as defined in \cite{Bruno:2014ova}, see \fig{f:taui}, which we use 
as a rough estimate of the exponential
autocorrelation time $\tau_{\rm exp}$, cf. \cite{algo:csd}.
            
At the smallest lattice spacing $a=0.036\,\fm$ 
that we reach with standard Wilson fermions
we estimate $\tau_{\rm exp}\simeq200-300$ MDU (Molecular Dynamics Units).
Our statistics of $4000-8000$ MDU is therefore adequate but does require
a particularly careful error analysis.
With twisted mass fermions at maximal twist we reach a smallest lattice
spacing of $a=0.023\,\fm$ ($\beta=6.0$). There we estimate 
$\tau_{\rm exp}=357$ MDU and have a statistics of $63\tau_{\rm exp}$.
  For the twisted mass simulations at $M/\Lambda=4.87$ and $\beta=5.88\,,\;6.0$
  the statistics is too small to determine $\tau_{\rm int}$ for $\chi^\mathrm{corr}$.
The autocorrelation times shown in \fig{f:taui} are reasonably well described by
the dotted line
\begin{equation}\label{e:tauexp}
  \tau_{\rm exp} = 20 t_0/a^2 \,,
\end{equation}
where one has to take into account that determinations of $\tau_{\rm exp}$
including an error estimate
are notoriously difficult.
Thus the data in \fig{f:taui} is consistent with the expectation, that for 
simulations with open boundary conditions autocorrelation times scale 
with $1/a^2$.

The error analysis is performed with the program\footnote{{\tt http://www-zeuthen.desy.de/alpha/}}
of \cite{algo:csd}. It is based on \cite{Wolff:2003sm} and adds a tail to the 
autocorrelation function as an estimate of the slow mode
contribution~\cite{algo:csd}.

\section{Non-perturbative mass dependence} \label{s:np}

\subsection{Test of the factorization formula}
\label{s:np-factorization}

We remind that our model is QCD with two heavy, mass-degenerate quarks and thus the effective
theory, \qcdm, is the Yang-Mills theory up to $1/M^2$ corrections ($\nq=2$, $\nl=0$).
For the hadronic scale $\mscale=1/\sqrt{t_0}$ \cite{flow:ML}, the
factorization formula \eq{e:theequ} takes the form
\begin{equation}
  \label{e:theequroott0}
  \sqrt{\frac{t_0(M)}{t_0(0)}} =  \frac{1}{Q^{1/\sqrt{t_0}}_{0,2} \times P_{0,2}(M/\Lambda)} + \rmO((\Lambda/M)^2)
\end{equation}
with $Q^{1/\sqrt{t_0}}_{0,2}=
[\Lambda \sqrt{t_0(0)}]_{\nq=2} / [\Lambda \sqrt{t_0}]_{\nq=0}$.
We turn now to a comparison of \eq{e:theequroott0} to non-perturbative data.
Preliminary results have been presented in \cite{Knechtli:2015lux}, where
only data for Wilson fermions were available. Now we can combine those data
with the new simulations with twisted mass fermions and perform careful
continuum extrapolations. In the extrapolations we only use data points which
satisfy $a^2/t_0(M)<0.32$.

In order to compute the ratio in \eq{e:theequroott0} we write
\begin{equation}\label{e:theequroott0_via_L1}
  \sqrt{\frac{t_0(M)}{t_0(0)}} =
  \frac{\sqrt{t_0(M)}}{L_1} \times \left( \frac{\sqrt{t_0(0)}}{L_1} \right)^{-1}
\end{equation}
 and separately take continuum limits for the two factors on the right hand
side. There the mass independent scale $L_1$ enters, see \sect{s:simulation}. The pairs
$(L_1/a\,,\,\beta)$ are computed from a quadratic fit
of $\ln(L_1/a)$ as a function of $\beta$.
We take data for $L_1/a$ from table 13 of \cite{alpha:lambdanf2} and
add the newly determined values 
$L_1/a=20.31(69)$ at $\beta=6.1569$ and
$L_1/a=24.83(88)$ at $\beta=6.2483$.
%%%%%%%%%%%%%%%%%%%%%%%%%%%%%%%%%%%%%%%%%%%%%%%%%%%%%%%%%%%%%%%%%%%%%%%%%%%%%%%
\begin{figure}[t]
\begin{center}
   \includegraphics[width=0.49\textwidth]{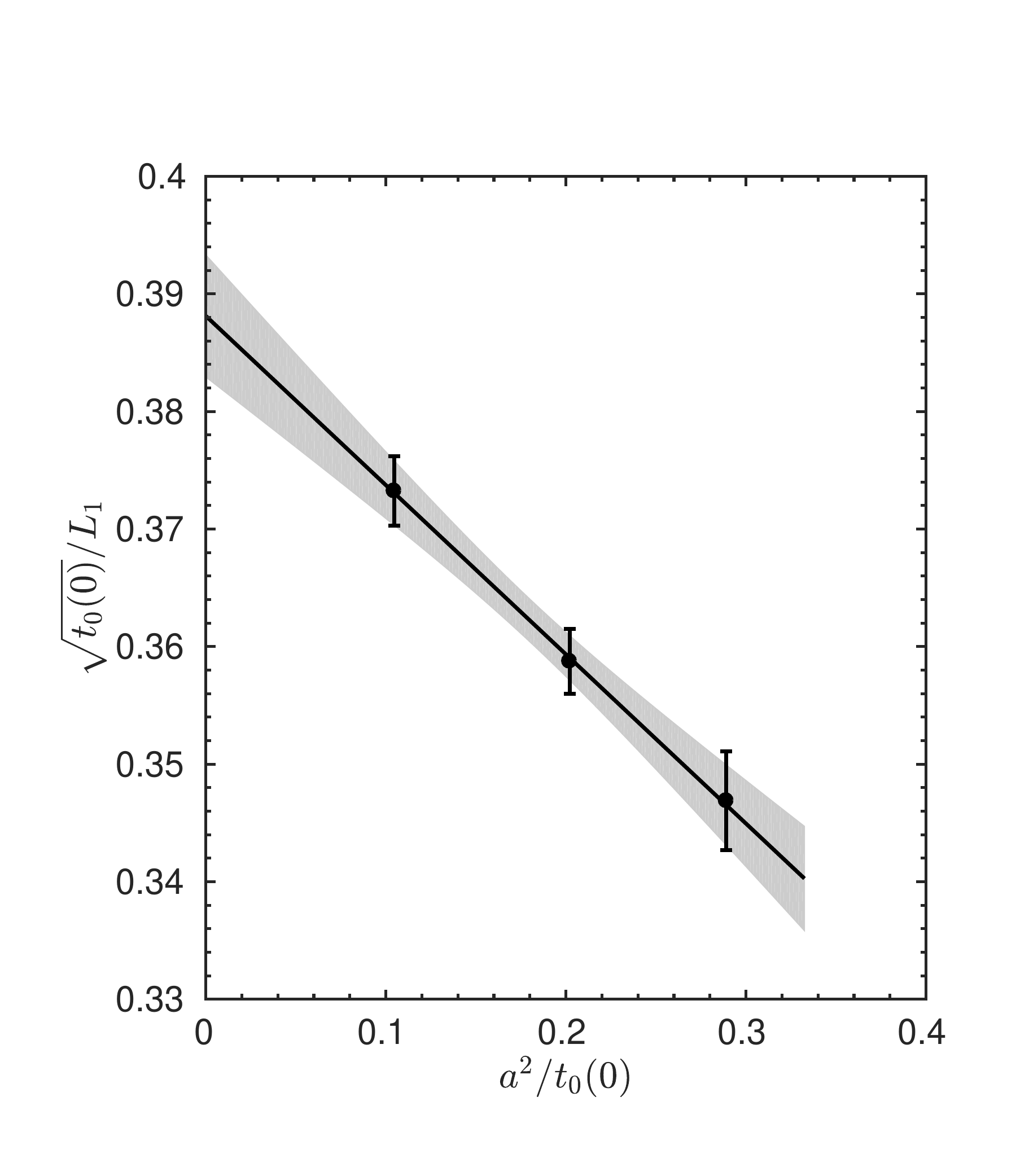}
   \includegraphics[width=0.49\textwidth]{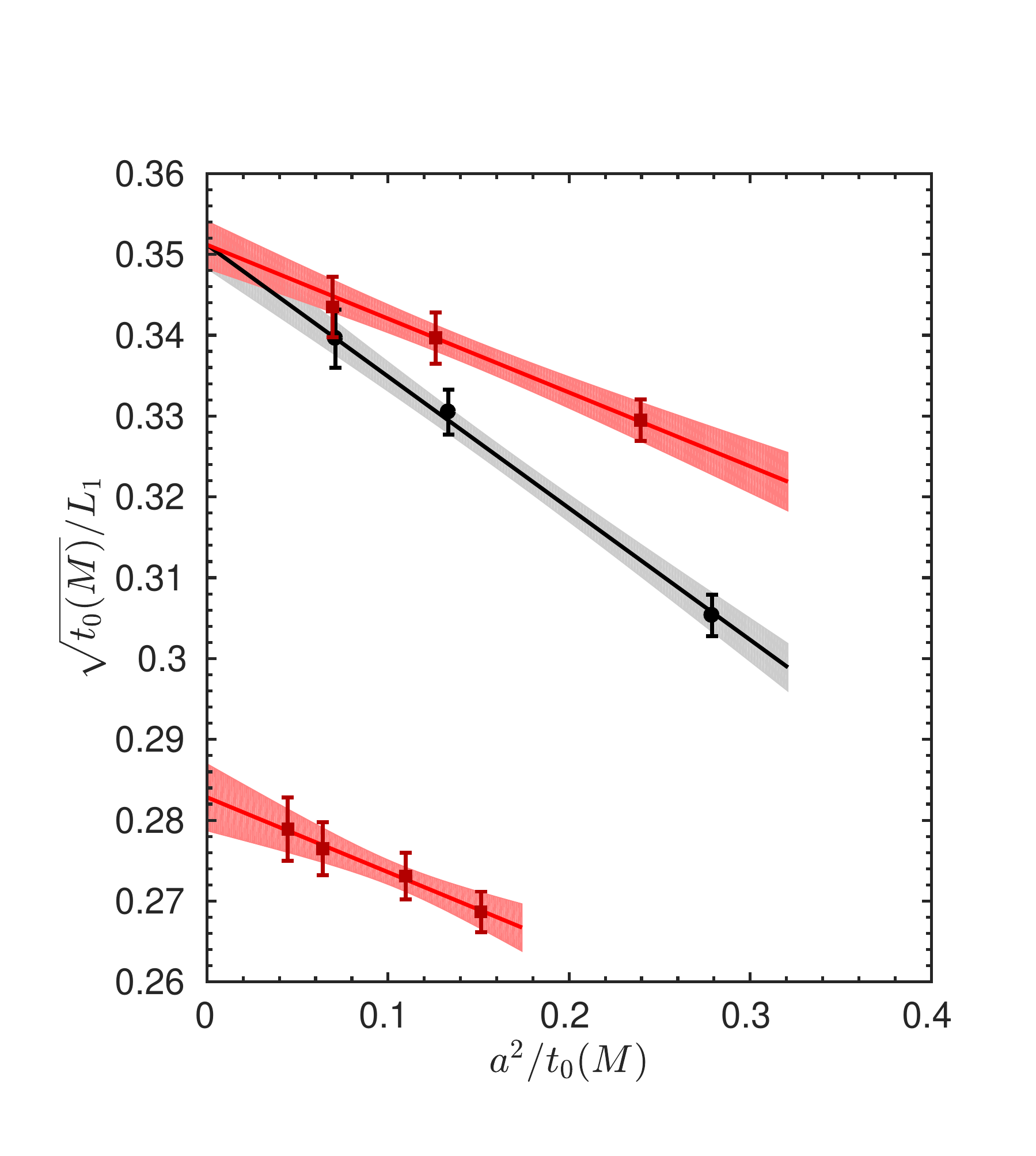}
\end{center}
\caption{Continuum extrapolations of $\sqrt{t_0}/L_1$,
  using a linear extrapolation in $a^2/t_0$.
  The shaded bands are the extrapolation errors.
  The left plot shows the results for $\sqrt{t_0(0)}$
  in the chiral limit.
  The right plot shows the results for $\sqrt{t_0(M)}$ 
  at $M/\Lambda=0.59$ (upper data set in the plot) and
  $M/\Lambda=4.87$ (the charm quark mass $\Mc$, lower data set in the plot).
  Black circles represent the standard Wilson and
  red squares the twisted mass discretizations.
  Where both are available a combined continuum extrapolation is
  performed.
  }
\label{f:cl_bigplot}
\end{figure}
%%%%%%%%%%%%%%%%%%%%%%%%%%%%%%%%%%%%%%%%%%%%%%%%%%%%%%%%%%%%%%%%%%%%%%%%%%%%%%%

%%%%%%%%%%%%%%
\begin{table}[t]
  \centering
  \begin{tabular}{cc}
    \toprule
    $M/\Lambda$ & $\sqrt{t_0(M) / t_0(0)}$ \\
    \midrule
    0.5900 & 0.9048(\phantom{0}43) \\
1.2800 & 0.8458(\phantom{0}74) \\
2.5000 & 0.7880(\phantom{0}73) \\
4.8700 & 0.7287(127) \\
5.7781 & 0.7151(102) \\
\bottomrule
\end{tabular}
  \caption{The values of $\sqrt{t_0(M) / t_0(0)}$ computed through
      \eq{e:theequroott0_via_L1}. The errors are obtained from error
      propagation which takes into account the correlation between the
      two factors in \eq{e:theequroott0_via_L1}.
    }
    \label{t:sqrt_t0M_over_t0c}
\end{table}
%%%%%%%%%%%%%%

%%%%%%%%%%%%%%%%%%%%%%%%%%%%%%%%%%%%%%%%%%%%%%%%%%%%%%%%%%%%%%%%%%%%%%%%%%%%%%%
\begin{figure}[t]
\begin{center}
   \includegraphics*[angle=0,width=.65\textwidth]{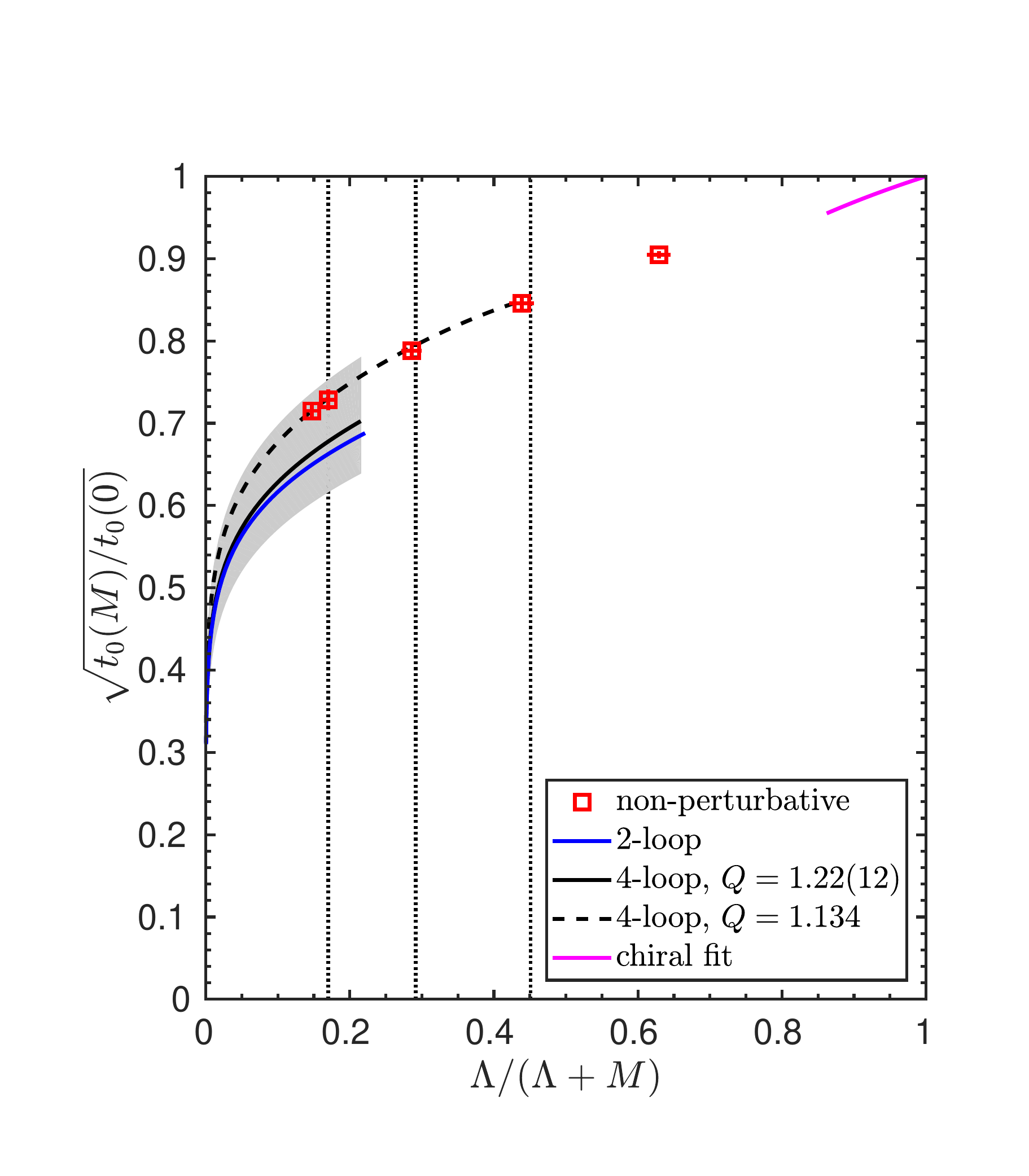}
\end{center}
\caption{The mass-dependence of the ratio $\sqrt{t_0(M)/t_0(0)}$ in
  the theory with two mass-degenerate quarks.
  Monte Carlo data after continuum extrapolation are compared with the
  perturbative predictions for $1/(QP)$ at large $M$ \eq{e:theequroott0}.
  The gray shaded error band represent the error of the 4-loop curve
  (black line) deriving from $Q$.
  The dashed line is the 4-loop curve adjusting the value of
  $Q$ to go through the point at $M/\Lambda=5.7781$, see \eq{e:predictedQ}.
  A fit function which describes the mass-dependence close to the chiral
  limit is also shown to the right.
  The vertical dotted lines mark the values of the quark mass
  $\Mc$, $\Mc/2$ and $\Mc/4$.}
\label{f:bigplot}
\end{figure}
%%%%%%%%%%%%%%%%%%%%%%%%%%%%%%%%%%%%%%%%%%%%%%%%%%%%%%%%%%%%%%%%%%%%%%%%%%%%%%%

The first factor on the right hand side of \eq{e:theequroott0_via_L1}
is computed using the data $t_0(M)/a^2$ obtained in the simulations
listed in \tab{t:ens-Wilson} and \tab{t:ens-tm}.
For the simulations with standard Wilson fermions we
include the $\bg$ effects as explained in \sect{s:improvement}.
We have data for five values of the quark masses given in
\eq{e:targetM}.
Some of our data for the ratio $\sqrt{t_0(M)} / L_1$ are shown in
the right plot of \fig{f:cl_bigplot} together with their continuum
extrapolations.
We show the two extreme values of the quark mass, separated by
a factor of 8.
The extrapolations linear in $a^2/t_0$ work very well and we observe that
the size of cut-off effects is smaller for the twisted mass data.
For this reason we opted for the twisted mass discretization to simulate
masses at or larger than the charm quark mass.

In order to compute the second factor on the right hand side of
\eq{e:theequroott0_via_L1} we use
the values of $t_0(0)/a^2$ in the chiral limit which are known for 
$\beta=6/g_0^2=5.2$, $5.3$ and $5.5$ from \cite{Bruno:2013gha}.
The continuum extrapolation of $\sqrt{t_0}(0)/L_1$
linear in $a^2/t_0(0)$ using
the three $\beta$ values works well, see the left plot of
\fig{f:cl_bigplot} and yields $\sqrt{t_0(0)}/L_1=0.3881(52)$.

Our continuum results for the ratio $\sqrt{t_0(M) / t_0(0)}$
are listed in \tab{t:sqrt_t0M_over_t0c}. Correlations
of the two factors originating from the common data of the scale $L_1/a$
help to reduce the overall error.
    
\Fig{f:bigplot} shows the values of $\sqrt{t_0(M) / t_0(0)}$
of \tab{t:sqrt_t0M_over_t0c} as a function of $\Lambda/(\Lambda+M)$.
We display a horizontal error stemming from the uncertainty of
$M/\Lambda$ originating from $\Lambda L_1$ in \eq{e:MoL}.
The vertical dotted lines mark the values of the quark mass
$\Mc$, $\Mc/2$ and $\Mc/4$.
We compare the Monte Carlo data to the factorization formula
\eq{e:theequroott0},
where the factor $P_{0,2}$ is computed to 2- (blue dashed line) and 4-loops 
(black line). 
The error on the factorization formula
comes from the
 numerical values
$[\Lambda L_1]_{\nq=2} = 0.629(36)$ \cite{alpha:lambdanf2},
$[\sqrt{t_0}/{L_1}]_{\nq=2} = 0.3881(52)$,
$[\Lambda r_0]_{\nq=0} = 0.602(48)$ \cite{mbar:pap1},
$ [\sqrt{t_0}/r_0]_{\nq=0}= 0.3319(19)$ \cite{Knechtli:2017xgy}
combined to
\begin{equation}\label{e:Q}
  Q^{1/\sqrt{t_0}}_{0,2} =
  \frac{[\Lambda L_1]_{\nq=2} \times [\sqrt{t_0(0)}/{L_1}]_{\nq=2}}
  {[\Lambda r_0]_{\nq=0} \times [\sqrt{t_0}/r_0]_{\nq=0}}
  = 1.22(12)
\end{equation}
and is 
displayed by the gray shaded band only for the 4-loop curve.
For completeness, in \fig{f:bigplot} the magenta line to the right shows
the mass dependence in the chiral limit estimated from~
\cite{Sommer:2014mea,Bruno:2013gha},
cf. \cite{Bar:2013ora}.

From \fig{f:bigplot} we see that there is agreement between the
Monte Carlo data of \tab{t:sqrt_t0M_over_t0c}
and the factorization formula \eq{e:theequroott0}
for quark masses at the charm quark mass value $\Mc$. 
Thus within our 
precision of 10\% due to the uncertainty of the factor $Q$ in
\eq{e:Q}, the data match the upper error band of the perturbative prediction.
In \cite{Bruno:2014ufa} we presented results
for the ratio $r_0(M)/r_0(0)$ and reached similar conclusions albeit
with less precise data covering only the region  below the charm quark mass. Our new results for  
$\sqrt{t_0(M) / t_0(0)}$ are much more precise than the 
value of $Q$ extracted from the literature.
This allows to turn the tables and predict
\bes\label{e:predictedQ}
    Q^{1/\sqrt{t_0}}_{0,2} = 1.134(28) \,,
\ees
obtained by taking $M/\Lambda=5.7781$
in \eq{e:theequroott0}. For $\sqrt{t_0(M)/t_0(0)}$ we use
our result in the last line of \tab{t:sqrt_t0M_over_t0c}.
We evaluate the factor $P_{0,2}(M/\Lambda=5.7781)=1.2328$
and assign to it a conservative 2\% error as it will be estimated in \sect{s:disc}.
This determination avoids entirely the computation of the running
of the coupling at high energy \cite{mbar:pap1,alpha:nf2}. 
In a nutshell it is replaced by perturbation theory for the difference of the running. The essential point is that the latter is given by the contribution of quark loops for which we 
non-perturbatively confirm that perturbation theory is very 
accurate.
We will comment more on this in the conclusions.

\subsection{The mass-scaling function $\etargi$}
\label{s:np-etaM}

By discretizing the derivative in \eq{e:etaM2} we obtain from our
simulations numerical estimates of the mass-scaling function 
\bes\label{e:etaMest}
     \etargi(\overline{M}) \approx {\log(\mscale(M_2)/\mscale(M_1)) \over \log(M_2/M_1)}\,,
         \quad \overline{M}=\sqrt{M_2 M_1} \,.
\ees
We use this definition to compute $\etargi(\overline{M})$ at
$\overline{M}=\sqrt{1.28\times0.59}$ and
$\sqrt{2.50\times1.28}$ using
$\mscale=1/\sqrt{t_0}$, $1/\sqrt{t_c}$ and $1/w_0$.
As emphasized before, these estimates
differ by $1/M^2$ effects.
We have data at three values of the lattice coupling
$\beta=6/g_0^2=5.3$, $5.5$ and $5.7$ for both
standard Wilson and twisted mass discretizations.
  We can also compute a value of $\etargi(\overline{M})$ at
  $\overline{M}=\sqrt{4.87\times5.7781}$ but its statistical errors are
  large.

%%%%%%%%%%%%%%%%%%%%%%%%%%%%%%%%%%%%%%%%%%%%%%%%%%%%%%%%%%%%%%%%%%%%%%%%%%%%%%%
\begin{figure}[t]
\includegraphics[width=0.49\textwidth]{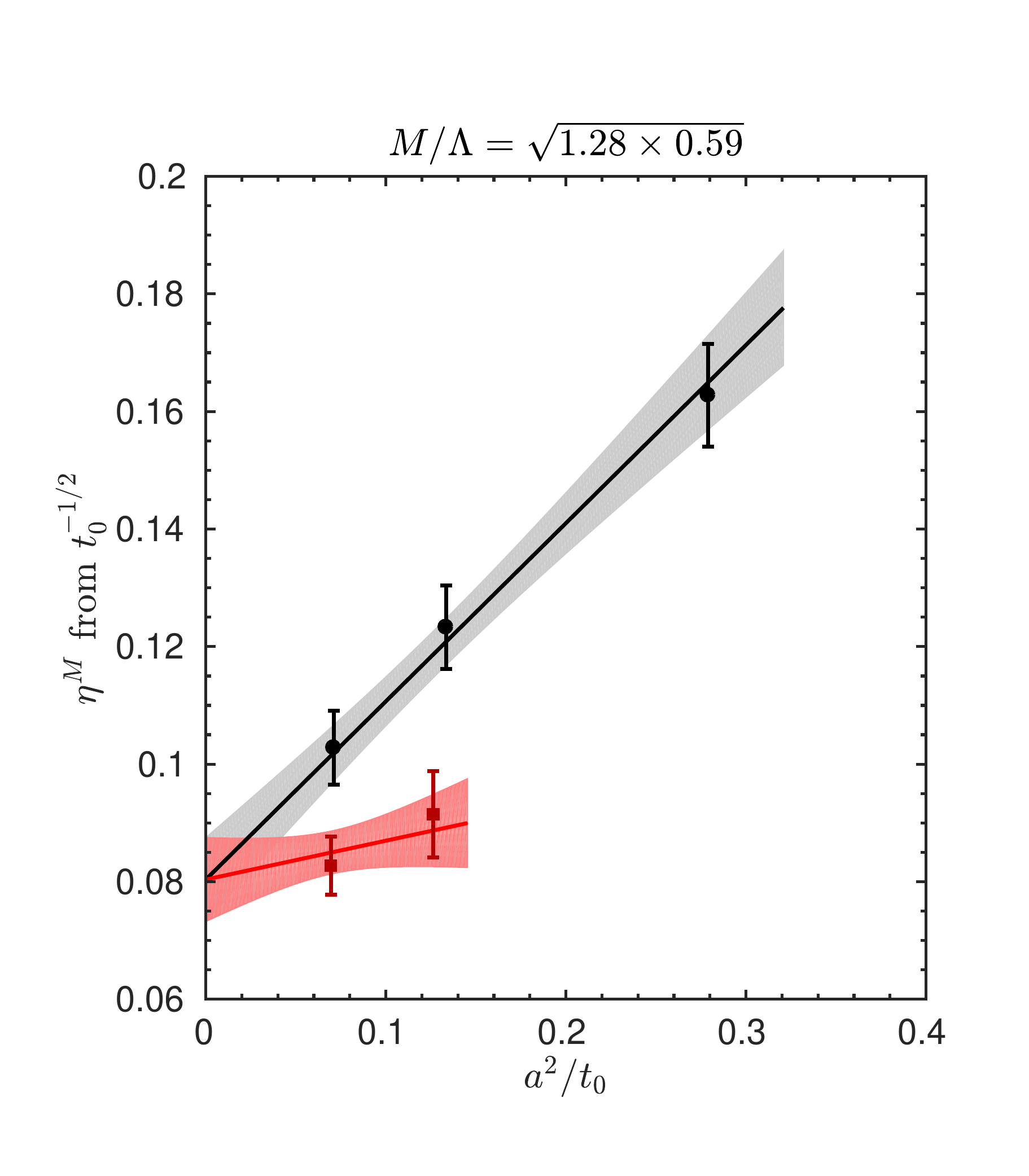}
\includegraphics[width=0.49\textwidth]{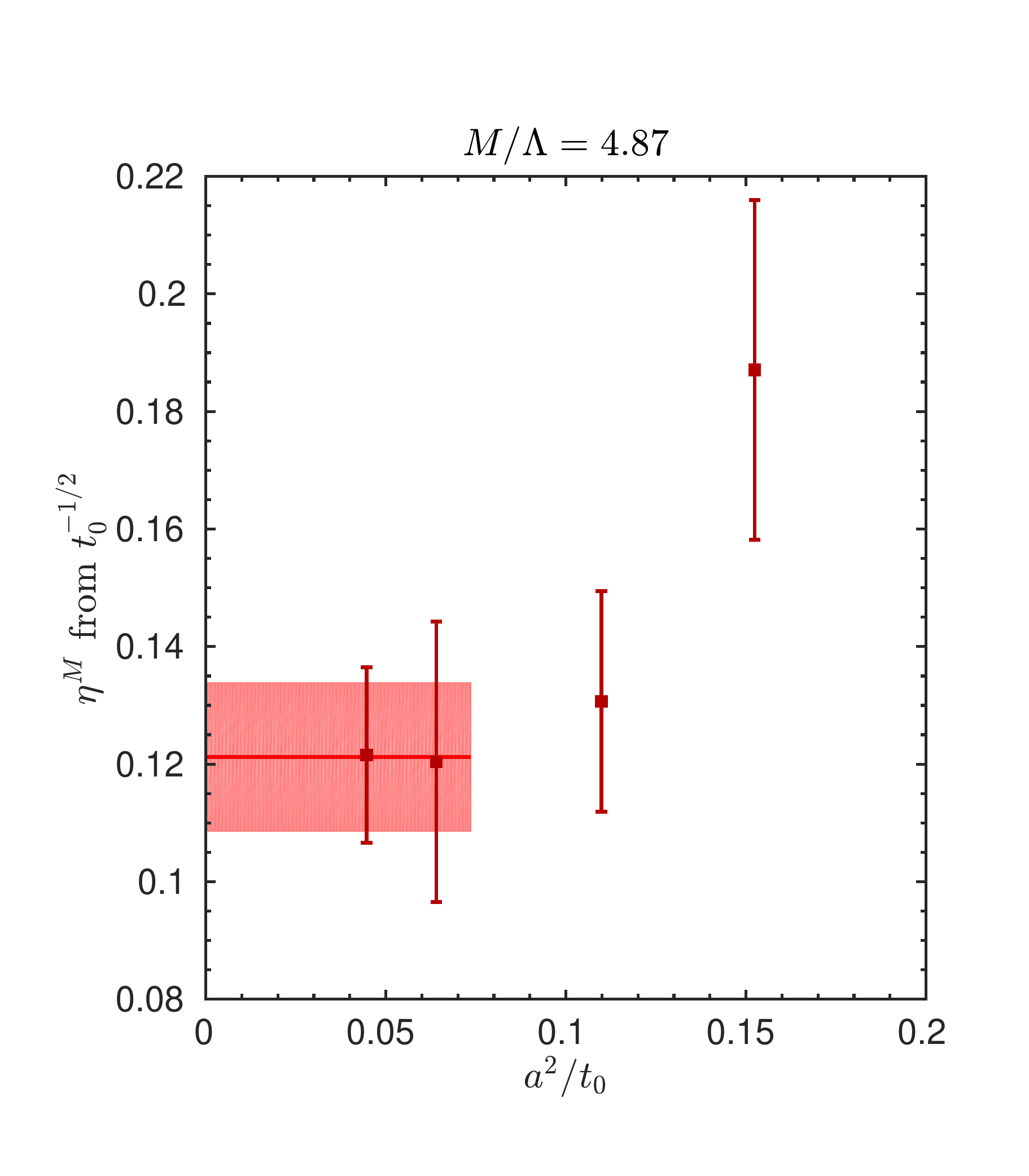}
\caption{Examples of continuum limits of $\etargi(M)$
  extracted from $\mscale=1/\sqrt{t_0}$ using a
  linear extrapolation in $a^2/t_0(M)$.
  In the left plot $\etargi(\overline{M})$ is computed at
  $\overline{M}/\Lambda=\sqrt{1.28\times0.59}$ using the definition
  \eq{e:etaMest}.
  Shown are data for standard Wilson (black circles) and
  twisted mass (red squares) and their combined continuum extrapolation.
  In the right plot $\etargi(M)$ is computed at $M=\Mc$ ($M/\Lambda=4.87$)
  using the definition \eq{e:etaMderivt0}.}
\label{f:etaMextrap}
\end{figure}
%%%%%%%%%%%%%%%%%%%%%%%%%%%%%%%%%%%%%%%%%%%%%%%%%%%%%%%%%%%%%%%%%%%%%%%%%%%%%%%

%%%%%%%%%%%%%%%%%%%%%%%%%%%%%%%%%%%%%%%%%%%%%%%%%%%%%%%%%%%%%%%%%%%%%%%%%%%%%%%
\begin{figure}[t]
\begin{center}
   \includegraphics*[angle=0,width=\textwidth]{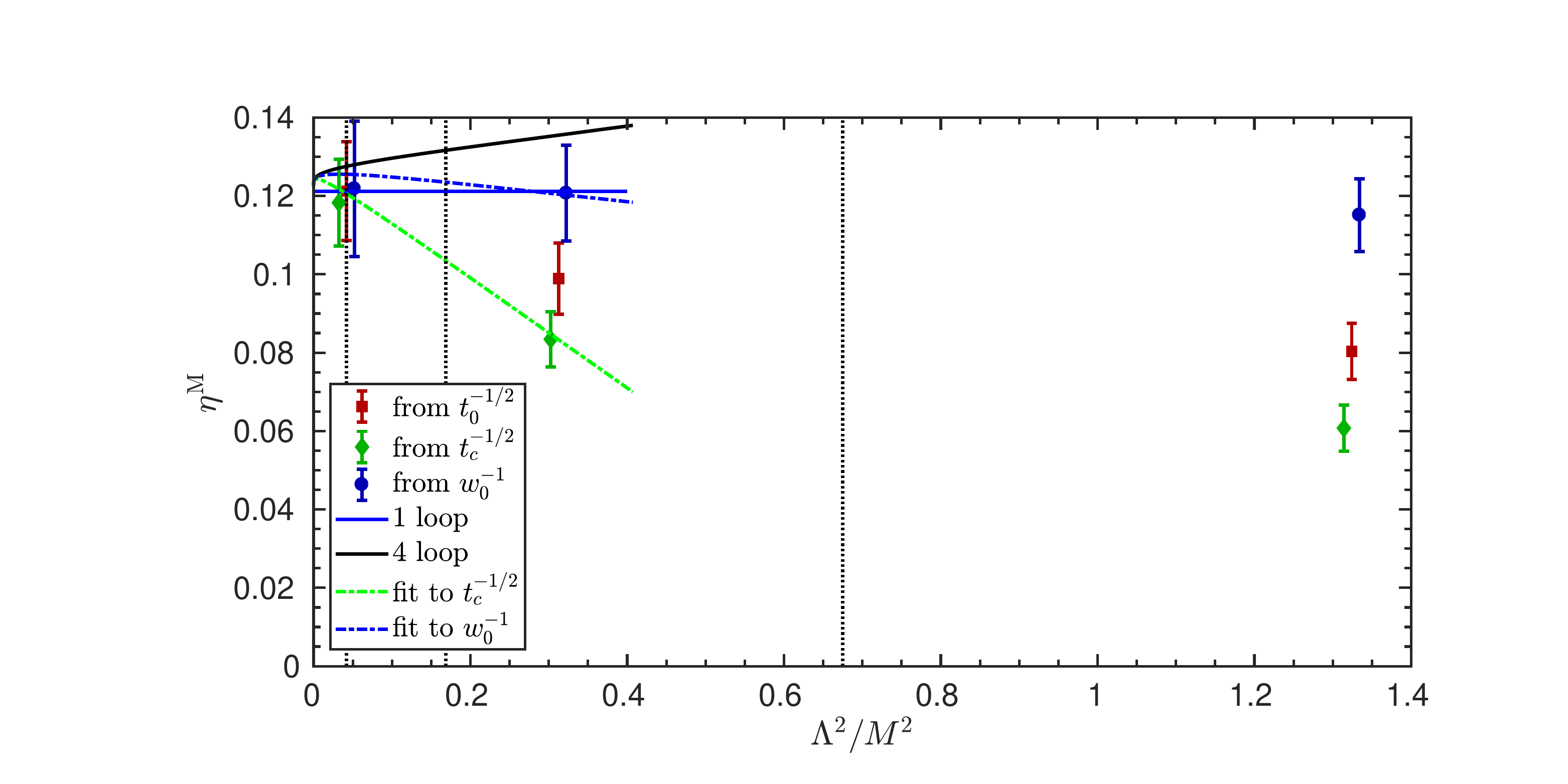}
\end{center}
  \caption{The mass dependence of the mass-scaling function $\etargi$ in
    the theory with two mass-degenerate quarks.
    $\etargi$ is obtained from
    the hadronic scales $1/\sqrt{t_0}$, $1/\sqrt{t_c}$ and $1/w_0$ and the
    data for a given mass $M$ are slightly diplaced horizontally for clarity.
    The Monte Carlo data are compared to the perturbative curves.
    The dash-dotted lines are the fits \eq{e:etaM_fit} and \eq{e:etaM_NP}
    for $1/\sqrt{t_c}$ and $1/w_0$.
    The vertical dotted lines mark the values of the quark mass
    $\Mc$, $\Mc/2$ and $\Mc/4$.
  }
\label{f:etaMall}
\end{figure}
%%%%%%%%%%%%%%%%%%%%%%%%%%%%%%%%%%%%%%%%%%%%%%%%%%%%%%%%%%%%%%%%%%%%%%%%%%%%%%%

For the case $\mscale=1/\sqrt{t_0}$ and $\overline{M}=\sqrt{1.28\times0.59}$,
the simulation data are shown in the left plot of \fig{f:etaMextrap}.
The continuum value results from a
combined continuum extrapolation linear in $a^2/t_0(M)$.
In all our continuum extrapolations we apply the cut $a^2/t_0(M)<0.32$ to
the data to be fitted.
The plot shows the continuum extrapolation
for both discretizations together with its error bands.

The continuum values of $\etargi(\overline{M})$
for the various choices of $\mscale$ are presented in
\fig{f:etaMall} and plotted against $\Lambda^2/\overline{M}^2$.
Notice that the data points corresponding to different quantities $\mscale$
are slightly displaced horizontally for clarity of presentation.
The spread
of the data due to $1/M^2$ effects decreases when
$\overline{M}$ increases as expected.
For comparison we plot in \fig{f:etaMall} also
the 1-loop (the constant value $\eta_0$) and 4-loop
(up to the $\etaM_3$ term) expressions, see \eq{e:exp-etaM}, \eq{e:etaMrec}
and appendix \ref{s:coefficients}.

The mass-scaling function $\etargi$ can also be computed directly
from a simulation at a single quark mass.
Using the twisted mass discretization we can rewrite \eq{e:etaM2},
for example taking $\mscale=1/\sqrt{t_0}$, as
\begin{equation}\label{e:etaMderivt0}
-\frac{\mu}{2t_0}\frac{\mathrm{d} t_0}{\mathrm{d} \mu}
=
\etargi(M) \,.
\end{equation}
The derivative $\frac{\mathrm{d}t_0}{\mathrm{d}\mu}$ 
is computed as explained in \sect{s:hadscales}.
Using $\mscale=1/\sqrt{t_c}$ or $1/w_0$ results in 
determinations of $\etargi(M)$ similar to \eq{e:etaMderivt0}.

In the right plot of \fig{f:etaMextrap} we show the data for the
quantity on the left-hand side of
\eq{e:etaMderivt0} computed from our
simulations at $M=\Mc$ ($M/\Lambda=4.87$)
with twisted mass fermions at four values of
the lattice coupling $\beta=6/g_0^2=5.6$, $5.7$, $5.88$ and
$6.0$. Our fine lattices are needed to control the cut-off effects at
this large value of the mass.
We perform continuum extrapolations by ``fits'' to a constant.
Taking three, two or just the last point yields 
results which are in agreement. We settle for the two-point average 
which of course has a larger error than the three-point one.
The continuum values are plotted in \fig{f:etaMall}, together with
similar determinations of $\etargi(\Mc)$ from
$\mscale=1/\sqrt{t_c}$ and $1/w_0$. At $M=\Mc$ the different determinations
agree well with each other signaling the smallness of the $1/M^2$
corrections \cite{Knechtli:2017xgy}.

For our model with two charm quarks
we see from \fig{f:etaMall} that $\etargi$ is about 1/10, both in perturbation theory and 
non-perturbatively. For a single charm quark there is an additional
factor 1/2. Thus 
a $2$\% shift of the charm quark mass leads only to a
$1$\textperthousand~change of a low energy hadronic quantity of mass-dimension one.

The precision of $\etargi(\Mc)$ that we can achieve
is around 10\%. Within this error the non-perturbative values agree with
the perturbative one. This does not look very precise,
but in absolute terms this is $\Delta\etargi=0.01$.
We put this into the perspective of phenomenology in the following
section.

\section{How big are the effects of charm loops?} \label{s:disc}

We recapitulate that the effects of charm loops at low energies
come in two classes. One is when we are concerned with dimensionless low energy
observables which do not refer to quantities at energies around or above
the charm mass. In lattice slang: the quantity is long distance and
the lattice spacing $a$ is set through long distance physics
in the theory with the heavy quark.
In this case the value
of the $\Lambda$-parameter drops out and the only effects of the
heavy quark mass are
due to the power corrections originating from $\lag{2}$ studied in
\cite{Bruno:2014ufa,Knechtli:2017xgy}. These effects are very small.
To be specific, when decoupling two charm quarks,
the power corrections in ratios of hadronic scales~\eq{e:hadscales}
were found to be approximately 0.4\%.

The prototype for the second class is given by the connection
of the fundamental scales of the four-flavor  and the three-flavor theory.
In our model it is the connection between the two-flavor theory and the zero-flavor
theory. The very relevant question is what the uncertainty is
when one uses
the perturbatively computed ${P_{\tl,\tq}(M/\Lamq)}$.
In \sect{s:ptaccuracy} we have seen that 3,4,5-loop corrections are very small. How big can non-perturbative effects be? The close agreement
of our non-perturbative $\etargi$ (\sect{s:np-etaM}) with
perturbation theory and the dashed curve in \fig{f:bigplot} with
the non-perturbative points shows that they
are small. We now put this into numbers,
estimating the non-perturbative effects to $\etargi$
and to ${P_{\tl,\tq}(M/\Lamq)}$ in our model calculation with $\nq=2$, $\nl=0$. As will become clear, these estimates are rough
and, depending on the assumptions made, can vary quite a bit. Still, their 
smallness can be quantified at a reasonable level.

\subsection{Non-perturbative effects on $\etargi$ and $P_{0,2}$}
\label{s:NPeffects}

In \fig{f:etaMall} we include
dash-dotted curves corresponding to the fits
\bes\label{e:etaM_fit}
 \etargi= \etargi_{\rm pert} + \eta^{\mathrm{M},\,\mscale}_{\rm NP}\,,
\ees
where $\etargi_{\rm pert}$
is the 4-loop expression and $\eta^{\mathrm{M},\,\mscale}_{\rm NP}$ the remainder, which depends
on the quantity $\mscale$.
As a first estimate of the non-perturbative contribution we assume that 
$\eta^{\mathrm{M},\,\mscale}_{\rm NP}$ is dominated 
by the terms in $\lag{2}$ and neglect the logarithmic (in $M/\Lambda$) corrections. This means we assume
\bes\label{e:etaM_NP}
    \eta^{\mathrm{M},\,\mscale}_\mathrm{NP} = c^{\mscale} {\Lambda^2 \over M^2}\,
\ees
for large masses.
Note that the fit function \eq{e:etaM_fit} has the correct
asymptotics
\bes\label{e:etaM_asymptotics}
\lim_{M\to\infty} \etargi = \eta_0 \,,
\ees
as guaranteed by asymptotic freedom in the form $\lim_{M\to\infty} \gstar=0$.
In \fig{f:etaMall} we compare fits for $\mscale=1/\sqrt{t_c}$ and
$\mscale=1/w_0$. The fits include the Monte Carlo data of $\etargi$ for
$M/\Lambda=4.87$ (the charm-quark mass) and $M/\Lambda=\sqrt{2.50\times1.28}= 1.8$. They
yield the values $c^{1/\sqrt{t_c}}=-0.167(22)$ and $c^{1/w_0}=-0.048(39)$.
In the following we will take $c=-0.2$, which is
a conservative choice accommodating both values and their errors.
Covering the end of the error bars
at the charm would require values of $|c|$ larger by a factor two to three.

We recall from \eq{e:etaM1} that the mass scaling function is defined as
  $\etargi\equiv \left.{\partial \log(P_{0,2}) \over \partial x}\right|_{\Lambda}$,
  with $x=\log(M/\Lambda)$.
The effect of the $\Lambda^2 \over M^2$ term on
${P_{0,2}(M/\Lambda)}$,
\bes
    \Delta \log ({P_{0,2}}) 
    &\equiv& \log\left[ {P_{0,2}(M/\Lambda)} \right] -
    \log\left[ \left.{P_{0,2}(M/\Lambda)}\right|_\mathrm{pert} \right]
    \label{e:logP_NP}\\
    &=& 
    - \int_{\log(M/\Lambda)}^\infty h(x) \rmd x\,,
    \quad \text{with} \;\;
    h(\log(M/\Lambda)  ) = \etargi-\etargi_\mathrm{pert} \,,
    \label{e:DeltaP2}
\ees
is easily evaluated. From
$h(x)=c\,\rme^{-2x}$ one has
\bes
    \Delta \log ({P_{0,2}})  =
    -\frac{c}{2}  {\Lambda^2 \over M^2}\,.
    \label{e:DeltaP3}
\ees
    Note that due to the asymptotics \eq{e:etaM_asymptotics}
      the contribution to \eq{e:logP_NP} from the integration limit at $\infty$
      cancels in the difference.
Inserting $c=-0.2$ and the approximate charm-quark mass value
${\Lambda^2 \over \Mc^2} \approx 1/25$ yields
$\Delta \log ({P_{0,2}}) = 0.004$.
This means a 0.4\% change (or better uncertainty) due to
non-perturbative effects of the described form and magnitude.
In other words a 0.4\% precision for perturbation theory in the
conversion of the $\Lambda$-parameter.
We consider this a good estimate, but it clearly depends on the
assumptions made. Therefore, we present a second, 
very conservative, estimate. 
%%%%%%%%%%%%%%%%%%%%%%%%%%%%%%%%%%%%%%%%%%%%%%%%%%%%%%%%%%%%%%%%%%%%%%%%%%%%%%%
\begin{figure}[t]
\begin{center}
   \includegraphics*[angle=0,width=\textwidth]{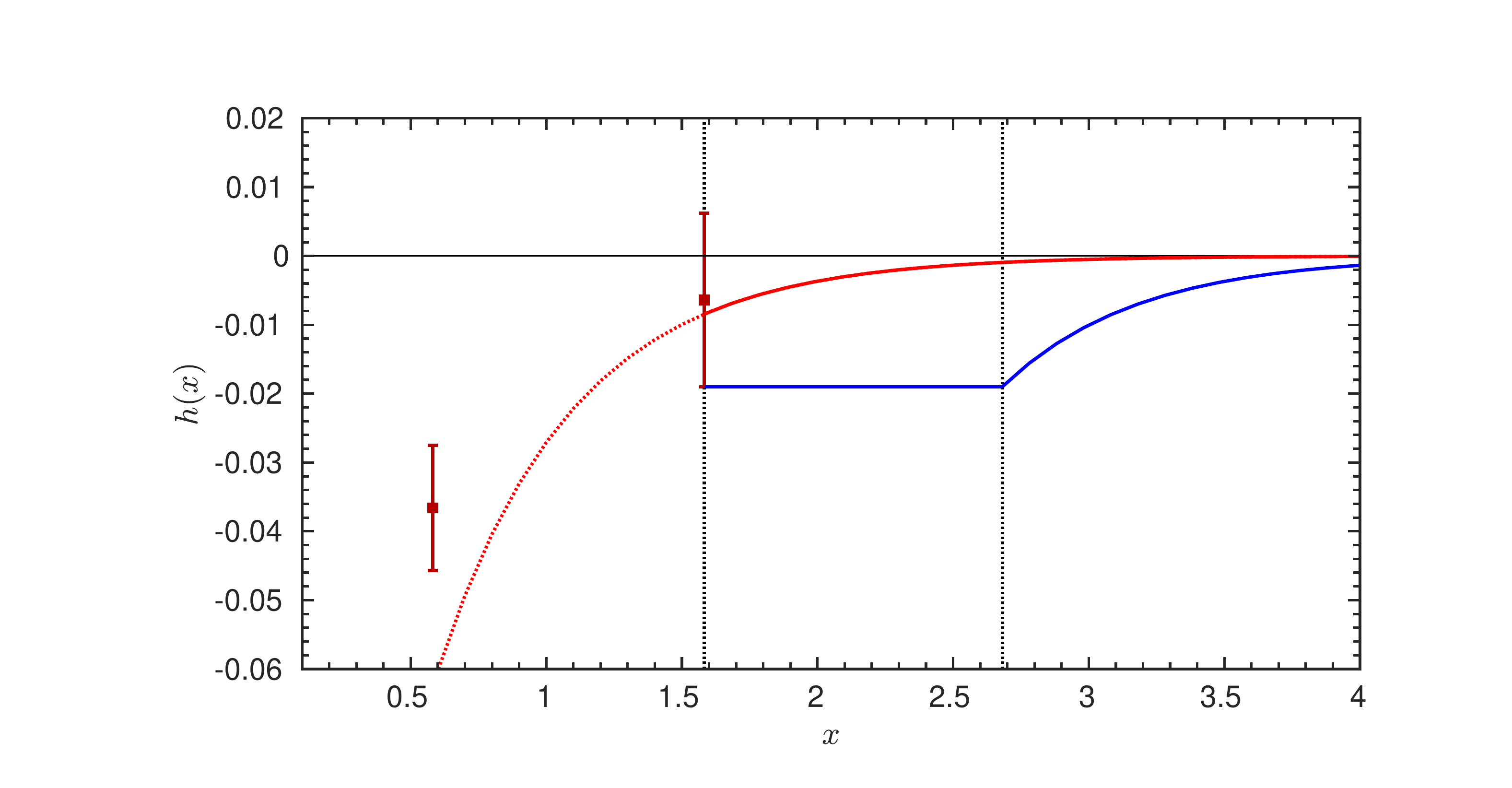}
\end{center}
\caption{The integrand  of \protect\eq{e:DeltaP2}
  for the scale $\mscale=1/\sqrt{t_0}$.
      Data points are $\etargi-\etargi_{\rm pert}$.  The 
      red line corresponds to the estimate \protect\eq{e:DeltaP3} and the blue line represents \protect\eq{e:DeltaP4} together with
      the $\exp(-2x)$ decay from $M=3\Mc$ on. 
    The vertical dotted lines mark the values of the quark mass
    $\Mc$ and $3\Mc$.
  }
\label{f:integrand}
\end{figure}
%%%%%%%%%%%%%%%%%%%%%%%%%%%%%%%%%%%%%%%%%%%%%%%%%%%%%%%%%%%%%%%%%%%%%%%%%%%%%%%

As illustrated in \fig{f:integrand}, we split the integral
into 
\bes
    \Delta \log ({P_{0,2}}) 
    &=& A + B\,,
    \\
    A&=& - \int_{\log(M/\Lambda)}^{\log(M_\mathrm{pert}/\Lambda)} 
      h(x) \rmd x\,,
      \\
    B&=& - \int_{\log(M_\mathrm{pert}/\Lambda)}^{\infty} 
      h(x) \rmd x\,,
\ees
where $M_\mathrm{pert}$ is high enough such that $h$ and therefore
$B$ can be neglected or replaced by the previous estimate.
For the lower mass region we just bound
\bes
    |A| &\leq & \log(M_\mathrm{pert}/M) \, h_\mathrm{max} \,,
    \label{e:DeltaP4}
\ees
where $h_\mathrm{max} $ is the maximum of $|h(x)|$ in the interval
$\log(M/\Lambda) \leq x \leq \log(M_\mathrm{pert}/\Lambda)$.
Numerical information is now obtained by making the reasonable assumption that beyond the masses that we have reached 
$\etargi$ continues approaching the perturbative one. We can then
replace $h_\mathrm{max}$ by what we find for our largest mass,
$\etargi(M_\charm/\Lambda)-\etargi_\mathrm{pert}(M_\charm/\Lambda)=-0.006(13)$ or $|h| \leq 0.019$. Further
setting
$M_\mathrm{pert} = 3 M_\charm$ where $1/M^2$ terms are suppressed
by an order of magnitude compared to at $M_\charm$,
we arrive at $|A| \leq  0.021$.
We here took the scale $\mscale=1/\sqrt{t_0}$ 
but the others 
yield numbers which are very close. 
Given that no decay of $|h|$ is used this is likely an
overestimate of the integral and we
neglect the small piece $B$. We thus cite as the 
conservative estimate
\bes\label{e:logP02_conservative}
    \Delta \log ({P_{0,2}}) = 0.02\,,
\ees
a 2\% non-perturbative contribution to $P_{0,2}$.

\subsection{Power corrections}
\label{s:powercorr}

%%%%%%%%%%%%%%%%%%%%%%%%%%%%%%%%%%%%%%%%%%%%%%%%%%%%%%%%%%%%%%%%%%%%%%%%%%%%%%%
\begin{figure}[t]
\begin{center}
   \includegraphics*[angle=0,width=\textwidth]{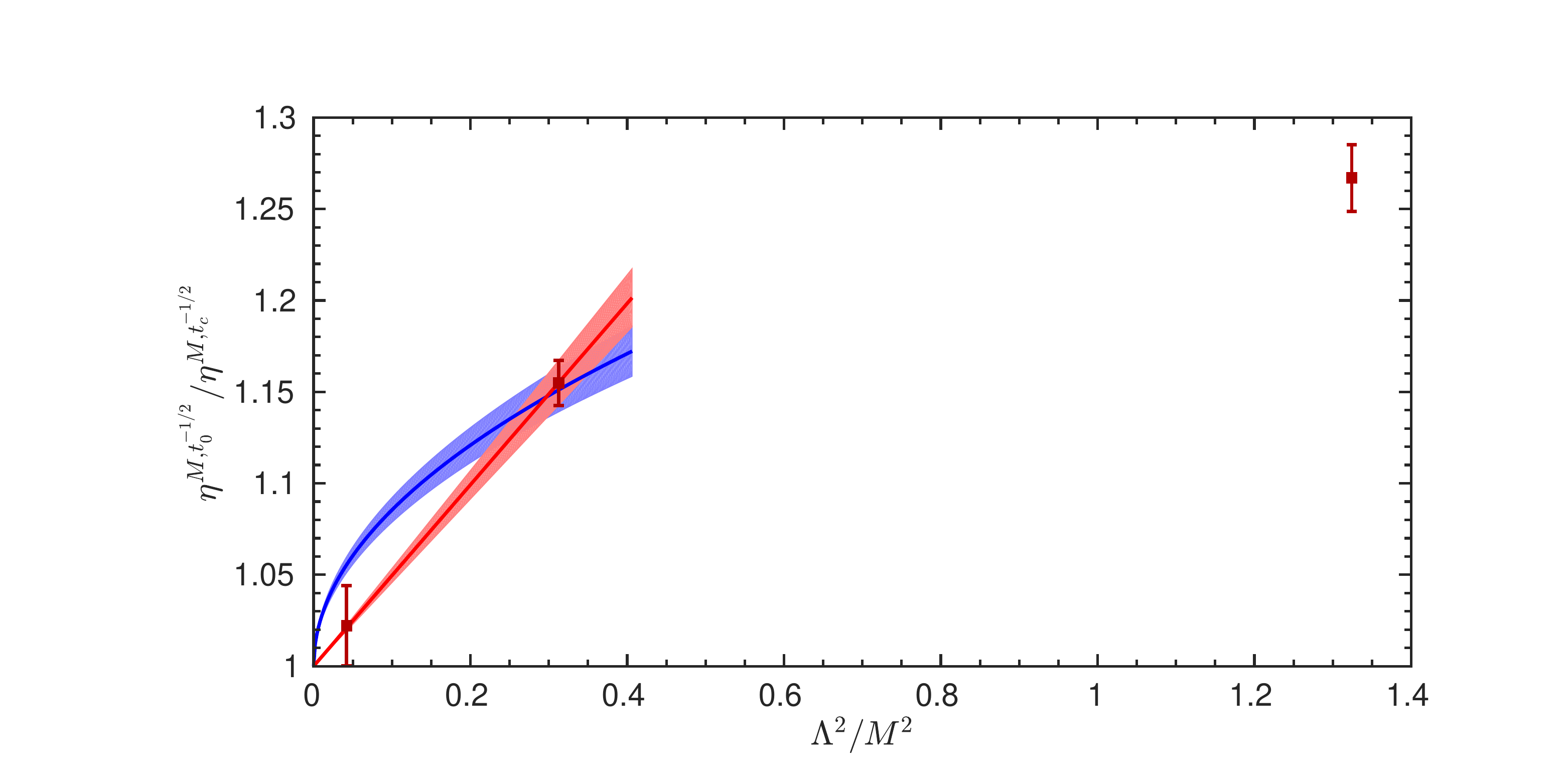}
\end{center}
\caption{The ratio \eq{e:etaratio} of the mass-dependence function $\etargi$
  computed from the hadronic scales $\mscale_1=1/\sqrt{t_0}$ and
  $\mscale_2=1/\sqrt{t_c}$.
  The lines in the red and blue bands are fits assuming leading non-perturbative
  effects proportional to $(\Lambda/M)^2$ and $\Lambda/M$ respectively.
  }
\label{f:etaratio}
\end{figure}
%%%%%%%%%%%%%%%%%%%%%%%%%%%%%%%%%%%%%%%%%%%%%%%%%%%%%%%%%%%%%%%%%%%%%%%%%%%%%%%

In \eq{e:etaM_NP} we made the
  assumption that the non-perturbative effects are dominated
  by the leading $(\Lambda/M)^2$ ones for our largest masses. It was tested in \cite{Knechtli:2017xgy}
  for ratios of two different hadronic scales $\mscale_1/\mscale_2$ in the same mass-range.
  We corroborate it for the case of $\etargi$ by computing the ratio
\bes\label{e:etaratio}
R = {\eta^{\mathrm{M},\,\mscale_1} \over \eta^{\mathrm{M},\,\mscale_2}}
\ees
of $\etargi$ calculated as in \eq{e:etaM2} from two different hadronic scales.
Using \eq{e:etaM_fit} and \eq{e:DeltaP2} we see that
$R = 1 + (h^{\mscale_1}-h^{\mscale_2})/\etargi_\mathrm{pert} + \rmO(h^2)$.
In \fig{f:etaratio} we show the results for the choice
$\mscale_1=1/\sqrt{t_0}$ and $\mscale_2=1/\sqrt{t_c}$.
The line in the red band is a fit to the two largest mass points using
the assumption in \eq{e:etaM_NP} and neglecting higher order terms in $R$.
It yields
$h^{1/\sqrt{t_0}}-h^{1/\sqrt{t_c}} = 0.50(4) \times (\Lambda/M)^2 \cdot \etargi_\mathrm{pert}$
with a $\chi^2$ per degree of freedom equal to 0.003.
For comparison we also show the line in the blue band
which corresponds to non-perturbative effects proportional to $\Lambda/M$.
It yields
$h^{1/\sqrt{t_0}}-h^{1/\sqrt{t_c}} = 0.27(2) \times \Lambda/M \cdot \etargi_\mathrm{pert}$ with
a worse $\chi^2$ per degree of freedom equal to $2.4$.

We can use the fits to the ratio $R$ to estimate the size of non-perturbative effects
in the difference of $\log(P_{0,2})$ extracted from $\mscale_1$ and $\mscale_2$:
\bes\label{e:logP02_difference}
\log\left[P_{0,2}^{\mscale_1}(M/\Lambda)\right] - \log\left[P_{0,2}^{\mscale_2}(M/\Lambda)\right]
= -\int_{\log(M/\Lambda)}^\infty \left[h^{\mscale_1}(x)-h^{\mscale_2}(x)\right] \rmd x\,.
\ees
Evaluating the integral with a constant
$\etargi_\mathrm{pert}\approx \etargi_\mathrm{pert}(\Mc)=0.1276$}
and ${\Lambda \over \Mc} \approx 1/5$ yields the values
$-0.0013$ (fit $h^{1/\sqrt{t_0}}-h^{1/\sqrt{t_c}} \sim (\Lambda/M)^2$) and
$-0.0069$ (fit $h^{1/\sqrt{t_0}}-h^{1/\sqrt{t_c}} \sim \Lambda/M$)
for the difference of $\log(P_{0,2})$ \eq{e:logP02_difference}.
This difference is a further test
of the non-perturbative effects. The absolute values are 
significantly smaller than the conservative estimate
in \eq{e:logP02_conservative}, confirming the latter.

\subsection{Heavy quark content of the nucleon}
\label{s:darkmatter}
The matrix element of the scalar heavy quark density between nucleon states is a
relevant contribution to the cross-section for the scalar interaction of dark matter with
ordinary matter \cite{Hisano:2017jmz}. It can be related, by the Hellmann--Feynman theorem,
to the derivative of the nucleon mass $m_N$ with respect to the heavy quark mass.
In the chiral limit for the up, down and strange quark and
up to $\rmO(\Lambda^2/M_q^2)$
this derivative is the mass-scaling function
$\etargi$, see \eq{e:etaM2},
\begin{equation}\label{eq:quarkcontent}
  \frac{1}{m_N} \langle N | m_{q,0} (\bar q q)_0 | N\rangle =
  \frac{1}{m_N}	\langle N | M_q (\bar q q)_{\rm RGI} | N\rangle =  \etargi + O(\Lambda^2 / M_q^2)\, ,
\end{equation}
where $m_{q,0}$ is the bare heavy quark mass and $(\bar q q)_0$ is the bare scalar density of quark $q$,
and $(\bar q q)_{\rm RGI}$ is the RGI-renormalized scalar heavy quark density.
Our result in \fig{f:etaMall} shows that perturbation theory can be safely applied
to compute $\etargi$ as it was done in \cite{Kryjevski:2003mh,Vecchi:2013iza,Ellis:2018dmb}
and non-perturbative effects in $\etargi$ are below $0.02/2$ for the case of
a single charm quark as just discussed.

\subsection{From the model to QCD}
\label{s:towardsQCD}

Note that currently the precision for the
$\Lambda$-parameter is at the level of around
4\% \cite{Aoki:2016frl,Olive:2016xmw}. This sets the scale for what is small and what is big.
Furthermore, there is no reason why our toy-model computation
should give a significantly different result for the magnitude
(not the details) of non-perturbative effects except that we have
decoupled {\em two} heavy quarks. Indeed, since we are dealing with small
effects of quark loops, it is very plausible that the effect
of more than one quark-loop effects are smaller than the ones
of a single quark loop, which scales proportionally to the number
of quarks. We are here just counting quark loops in arbitrary gauge backgrounds,
so the argument is valid independently of whether the gauge coupling
is large or small. It is  non-perturbative.
It means that these small effects will be about a factor two smaller
for the decoupling of the charm-quark in QCD, compared to the studied model.
We use this for the magnitude of all effects,
also for the uncertainty of perturbation theory.

We saw in \tab{t:PTcoeffs} that the dependence on the number of light quarks
of $\etargi$ between $\nl=0$ and $\nl=3$ amounts to about $20\%$ at leading order
in perturbation theory. For this reason we include a safety margin of $50\%$
in our estimate of non-perturbative effects $h_\mathrm{max}$, see \eq{e:DeltaP4}.
We conclude that one can {\em safely} neglect non-perturbative
effects all-together
for connecting three-flavor and four-flavor $\Lambda$
at a level down to
\bes\label{e:logP34_conservative}
    \Delta \log ({P_{3,4}}) = 0.015\,,
\ees
a 1.5\% non-perturbative contribution to $P_{3,4}$.

In the same way non-perturbative effects to \eq{eq:quarkcontent} are estimated to be
below $1.5\times h_\mathrm{max}/2=0.014$ in QCD when $q$ is the charm quark.

\section{Conclusions}
\label{s:concl}

In this article we presented a numerical study of the decoupling of heavy 
quarks. In particular we study the dependence of hadronic, low energy
quantities on the mass $M$ of the decoupled heavy quark.
We define and compute in perturbation theory a mass-dependence function
$\etargi$ \eq{e:etaM2}.
This computation is performed in leading order in the effective theory
which describes the decoupling of the heavy quarks at low energy.
We study the behavior of perturbation theory for the function $\etargi$
and show that perturbation theory by itself suggests that it is
well within the region of asymptotic convergence even for the
case of decoupling a charm quark. 
We remark that $\etargi$ can be related to the heavy quark content of
the nucleon, see \eq{eq:quarkcontent}, which is a relevant input for
dark matter searches.

To test the applicability of perturbation theory at the charm quark mass
we compare the mass dependence of the ratio $\sqrt{t_0(M)/t_0(0)}$ defined
in terms of the hadronic scale $1/\sqrt{t_0}$ to the perturbative prediction,
see \fig{f:bigplot}.
We also determine the mass-scaling function $\etargi$ non-perturbatively,
see \fig{f:etaMall}.
In order to be able to control the continuum extrapolations and have
precise results
we do this in a model consisting of two mass-degenerate quarks whose mass
ranges up to the charm quark mass.
The non-perturbative mass dependence agrees with the perturbative
prediction at a level of about 10\% for the small mass-scaling
function $\etargi$ computed at the charm quark mass.
This means that we confirm
  that a $2$\% shift of the charm quark mass leads
  only to a $1$\textperthousand~change of a low energy hadronic quantity of mass-dimension one.
We explained in \sect{s:disc} that this precision
is good enough to conclude that at the charm mass, the function $P_{\tl,\tq}$ in \eq{e:lamrat}
can be predicted by perturbation theory with
2\% accuracy for 
$\nq=2,\,\nl=0$ and 1.5\% accuracy for 
$\nq=4,\,\nl=3$. This allows to predict
\bes
    {\left.\Lambda_\msbar\sqrt{t_0(0)}\right|_{\nq=2} 
    \over
    \left.\Lambda_\msbar\sqrt{t_0}\right|_{\nl=0} }    
    = 1.134(28) \,.
\ees
Moreover we estimate that the non-perturbative effects in $\etargi$ are below 0.014 for
  the charm quark. These numbers are for the blue curve in \fig{f:integrand}, while we think that the red curve $\sim 1/M^2$ 
  is more realistic; it yields non-perturbative uncertainties which are a factor five smaller
  for $P_{\tl,\tq}$.

On the other hand, in the direct comparison of $\sqrt{t_0(M_c)/t_0(0)}$
to the product $Q\,P$, \eq{e:theequ} we presently have only 
10\% accuracy because in the literature the ratio, $Q$ is not known 
more precisely. 

Our most important conclusion concerns phenomenology:  
the ratio of three-flavor and four-flavor $\Lambda$-parameters 
can be computed in perturbation theory with a precision of 
1.5\% or better.
Power corrections $\sim 1/\Mc^2$ were found to be much smaller in low energy observables \cite{Bruno:2014ufa,Knechtli:2017xgy}. This means that the $\Lambda$-parameter of the five-flavor theory is safely predicted at the 1-2 percent level
from three-flavor low energy 
physics once the running of the coupling is under control~\cite{Bruno:2017gxd}, see \sect{s:towardsQCD} for details.
Note that the present precision of $\Delta \alpha_\msbar(M_\mathrm{Z}) = 0.0008$ of~\cite{Bruno:2017gxd} 
corresponds to 3.5\% in the $\Lambda$-parameter. Thus, there is plenty of room for relevant improvement within the three-flavour theory.

Similarly we conclude that non-perturbative effects to the charm quark content of the nucleon,
  \eq{eq:quarkcontent} are below 0.014.

{\bf Acknowledgement.}
We thank M.~Bruno and J.~Heitger for their inputs for our analyses.
We thank M.~Dalla~Brida and A.~Ramos for providing valuable feedback on the manuscript.
We gratefully acknowledge the computer resources
granted by the John von Neumann Institute for Computing (NIC)
and provided on the supercomputer JUROPA at J\"ulich
Supercomputing Centre (JSC) and by the Gauss Centre for
Supercomputing (GCS) through the NIC on the GCS share
of the supercomputer JUQUEEN at JSC,
with funding by the German Federal Ministry of Education and Research
(BMBF) and the German State Ministries for Research
of Baden-W\"urttemberg (MWK), Bayern (StMWFK) and
Nordrhein-Westfalen (MIWF). 
We are further grateful for
computer time allocated for our project
on the Konrad and Gottfried computers at 
the North-German Supercomputing Alliance HLRN,
on the CHEOPS, a scientific supercomputer
sponsored by the DFG of the regional computing centre of the University of
Cologne (RRZK),
the Stromboli cluster at the University of Wuppertal
and the PAX cluster at DESY, Zeuthen.
This work is supported by the Deutsche Forschungsgemeinschaft
in the SFB/TR~55 and is based on previous work \cite{Bruno:2014ufa}
supported also by the SFB/TR~09.
FK thanks CERN for hospitality.

\appendix

\section{Expansion of the matching condition and the mass scaling function}
\label{s:coefficients}
The coefficients of the matching of the coupling \eqref{e:matchg} can be found in
\cite{Grozin:2011nk,Kniehl:2006bg,Chetyrkin:2005ia}.
We collect here all known coefficients for convenience.  Note that we use the particular scale $\mu=\mstar$, for which logarithms $\log(\mu/\mbar(\mu))$ vanish and
$c_1=0$. The two loop coefficient is known for arbitrary $\nq\,,\nl$
\begin{equation}
c_2 = (\nq-\nl) \, {11\over 72}\,(4\pi^2)^{-2}\,, 
\end{equation}
The three loop one is known for $\nq-\nl=1,2$
\bes
c_3 &=&  
\big[1.881732 - 0.169303\,\nl\big]\,(4\pi^2)^{-3}
\quad \mbox{for }\nq-\nl=2\,,
\\
c_3 &=&  
\big[0.972057 - 0.084651 \,\nl\big]\,(4\pi^2)^{-3}
\quad \mbox{for }\nq-\nl=1\,,
\ees
and the four loop one only for $\nq-\nl=1$
\bes
c_4 &=&
\big[5.170347 - 1.009932 \nl - 0.021978 \,\nl^2\big]
\,(4\pi^2)^{-4} \quad \mbox{for }\nq-\nl=1\,.
\ees

The coefficients of the expansion of the mass scaling function \eqref{e:exp-etam}
are obtained by expanding \eqref{e:etam-matching}. Up to four loop they are given by
\bes
     \eta_0 & = & 1 -{b_0(\nq) \over b_0(\nl)} \,, \\ 
     \eta_1 & = & (\eta_0-1) \left[\br_1(\nq)-\br_1(\nl)\right] \,, \\
     \eta_2 & = & (\eta_0-1) \left[ c_2 + \br_2(\nq) - \br_2(\nl)\right] - \br_1(\nl) \eta_1\\
\eta_3 & = & (\eta_0-1) \left[ 2c_3 + \br_3(\nq) - \br_3(\nl)\right]
- \br_1(\nl) \eta_2 + \left(c_2 - \br_2(\nl)\right)\eta_1 \\
\eta_4 & = & (\eta_0-1) \left[3c_4 + \br_4(\nq) - \br_4(\nl) + \br_1(\nq)c_3
- c_2\left(4c_2+\br_2(\nl)\right)\right] \nonumber\\
&   & - \br_1(\nl) \eta_3 + \left(c_2 - \br_2(\nl)\right)\eta_2
+ \left(c_3 - \br_3(\nl)\right)\eta_1 \,.
\ees
The evaluation of the coefficients require the knowledge of the
$\beta$-function of the coupling up to five loops \cite{vanRitbergen:1997va,Czakon:2004bu,Baikov:2016tgj,Luthe:2016ima,Herzog:2017ohr}.

The coefficients of the function \eqref{e:exp-etaM} are straightforwardly obtained from \eqref{e:etaMrec}. Their evaluation requires in addition the anomalous dimension up to four loops
\cite{MS:4loop2,MS:4loop3}.

\section{Asymptotic expression for $P(M/\Lambda)$}
\label{s:asymptotic}
In this section we derive the asymptotic expression \eq{e:P}. Starting point is the definition of
$P(M/\Lambda)$ as the ratio of the $\Lambda$-parameters. 
We are interested in the asymptotic behavior at large $M/\Lambda$. Since our matching/renormalization scale $\mu=m_*$ is tied to the mass $\mbar(m_*)=m_*$, large $M/\Lambda$ means small $\gstar=\gbar(m_*)$, cf.~section \ref{s:PT}. Therefore we neglect terms $\rmO(\gstar^2)$. Using \eq{e_lambdapar} one obtains (see also \eq{e:PLambda})
\bes
\log [P(M/\Lambda)] & = & \log(\Laml/m_*) - \log(\Lamq/m_*)\,\\
                    & = & I_g^{\tl}(\gstar\,\Cs(\gstar)) -  I_g^{\tq}(\gstar) \,,\\
					& = & \frac{\eta_0}{2b_0(\nq)\gstar^2} - \frac{b_1(\nl)}{2b_0(\nl)^2} \log(b_0(\nl)\gstar^2)\label{e:logPLambda1}\\ 
					  &&   + \frac{b_1(\nq)}{2b_0(\nq)^2} \log(b_0(\nq)\gstar^2)+ \rmO(\gstar^2)\,.\label{e:logPLambda2}
\ees
In order to replace the coupling we extract the asymptotic relation
between $\gstar$ and $M/\Lambda$ from \eq{e:gstarM}. Using the shorthands $\logML=\log(M/\Lambda)$ and $x=2b_0(\nq)\gstar^2$ the relation up to $\rmO(\gstar^2)$ is
\bes
	\logML = \frac{1}{x} -\frac{d_0}{2b_0(\nq)}\log(x) + \frac{b_1(\nq)}{2b_0(\nq)^2}\log(x/2) + \rmO(x) \,.\label{e:lvsgbar}
\ees
Taking the logarithm on both sides yields $\log(\logML)=-\log(x)+\rmO(x\log(x))$. Inverting gives the result
\bes
	\frac{1}{x} = \logML + \frac{d_0}{2b_0(\nq)}\log(\logML) - \frac{b_1(\nq)}{2b_0(\nq)^2}\log(\logML/2) + \rmO\left(\frac{\log(\logML)}{\logML}\right) \,.
\ees
Using these relations $\gstar$ can be eliminated from \eqref{e:logPLambda1}-\eqref{e:logPLambda2} and one arrives
at \eq{e:P}.

\section{Simulation parameters} \label{s:tables}

\Tab{t:ens-Wilson} and \tab{t:ens-tm} summarize the parameters of our
simulations of $\Nf=2$ mass-degenerate quarks using
O($a$) improved standard Wilson fermions and
twisted mass Wilson fermions at maximal twist respectively.

%%%%%%%%%%%%%%
\begin{table}[h!]
 \centering
{\small
\begin{tabular}{c c c c c c c c c}
  \toprule
  $\frac{T}{a}\times\left(\frac{L}{a}\right)^3$ &  $\beta$ & BC &
  $\kappa$ & $am$ & $M/\Lambda$ & $r_0/a$   & $t_0/a^2$ & kMDU \\
\midrule
$64\times 32^3$  & $5.3$ & p &  0.13550 & 0.03405(8) & 0.638(46) &
5.903(36) & 3.481(14) & 1 \\
$64\times 32^3$  & $5.3$ & p &  0.13450 & 0.06979(7) & 1.308(95) &
5.193(20) & 2.714(14) & 2 \\
$64\times 32^3$  & $5.3$ & p &  0.13270 & 0.13873(8) & 2.600(189)  &
4.270(6)  & 1.842(3)  & 2 \\
\midrule
$120\times 32^3$ & $5.5$ & o &  0.136020 & 0.02467(4) & 0.630(46) &
8.49(12)  & 7.318(36) & 8 \\ 
$120\times 32^3$ & $5.5$ & o &  0.135236 & 0.05022(3) & 1.282(93) &
7.580(44) & 6.092(21) & 8 \\ 
$96\times 48^3$  & $5.5$ & p &  0.133830 & 0.09614(2) & 2.454(178) &
6.787(19) & 4.867(12) & 4 \\
\midrule
$192\times 48^3$ & $5.7$ & o &  0.136200 & 0.01691(2) & 0.586(43) &
11.48(24) & 14.02(6)  & 4 \\
$192\times 48^3$ & $5.7$ & o &  0.135570 & 0.03683(2) & 1.277(94) &
10.53(12) & 11.87(7)  & 4 \\
$192\times 48^3$ & $5.7$ & o &  0.134450 & 0.07209(2) & 2.500(184) &
9.50(5)   & 9.821(36) & 8 \\
\bottomrule
\end{tabular}
}
 \caption{Overview of the ensembles generated with 
   $\Nf=2$ O($a$) improved Wilson fermions.
   The columns show the lattice sizes,
   the gauge coupling $\beta=6/{g_0^2}$,
   the boundary conditions (periodic (p) or open (o)),
   the hopping parameter $\kappa$ (which is related to the bare mass
   $m_0$ through $\kappa=1/(2am_0+8)$),
   the PCAC mass $am$,
   the ratio of the RGI mass $M$ to the $\Lambda$ parameter (computed using \eq{e:MoL}),
   the scales $r_0/a$ and $t_0/a^2$ and
   the total statistics in molecular dynamics units.
}
 \label{t:ens-Wilson}
\end{table}
%%%%%%%%%%%%%%

\begin{table}[h!]
\centering
{\small
\begin{tabular}{c c c c c c c c}
\toprule
$\frac{T}{a}\times\left(\frac{L}{a}\right)^3$ &  $\beta$  & $\kappa$    & $a \mu$            & $M/\Lambda$ & $r_0/a$   & $t_0/a^2$ & kMDU \\
\midrule
$120\times 32^3$                              &  5.300    & 0.136457    & 0.024505           & 0.5900      &   --      & 4.174(13) & 4.3 \\
$120\times 32^3$                              &  5.500    & 0.1367749   & 0.018334           & 0.5900      & 8.77(15)  & 7.917(82) & 8 \\
$192\times 48^3$                              &  5.700    & 0.136687    & 0.013713           & 0.5900      &   --      & 14.40(10) & 5.8 \\
\midrule
$120\times 32^3$                              &  5.500    & 0.1367749   & 0.039776           & 1.2800      & 8.010(62) & 6.871(33) & 8 \\
$192\times 48^3$                              &  5.700    & 0.136687    & 0.029751           & 1.2800      &   --      & 12.668(39)& 16.2\\ 
\midrule
$120\times 32^3$                              &  5.500    & 0.1367749   & 0.077687           & 2.5000      & 7.392(62) & 5.836(27) & 8 \\
$192\times 48^3$                              &  5.700    & 0.136687    & 0.058108           & 2.5000      &   --      & 10.916(38)& 9 \\
\midrule
$192\times 48^3$                              &  5.600    & 0.136710    & 0.130949           & 4.8700      &   --      & 6.561(12) & 16 \\
$120\times 32^3$                              &  5.700    & 0.136698    & 0.113200           & 4.8703      & 9.123(57) & 9.104(36) & 17.2 \\
$192\times48^3$                               &  5.880    & 0.136509    & 0.087626           & 4.8700      & 11.946(55) & 15.622(62)& 23.1 \\
$192\times 48^3$                              &  6.000    & 0.136335    & 0.072557           & 4.8700      & 14.34(10) &22.39(12)& 22.4 \\
\midrule
$192\times 48^3$                              &  5.600    & 0.136710    & 0.155367           & 5.7781      &   --      & 6.181(11)& 2.1 \\
$192\times 48^3$                              &  5.700    & 0.136687    & 0.1343             & 5.7781      &   --      & 8.565(31) &  2.7 \\
$120\times 32^3$                              &  5.880    &  0.136509   & 0.103965           & 5.7781      &   --      &14.916(93)  &  59.9\\
\bottomrule
\end{tabular}
}
\caption{Overview of the ensembles generated with 
  $\Nf=2$ twisted mass fermions at maximal twist.
The columns show the lattice sizes,
the gauge coupling $\beta=6/{g_0^2}$, 
the hopping parameter $\kappa$ (for maximal twist), 
the twisted mass parameter $a\mu$,
the ratio of the RGI mass $M$ to the $\Lambda$ parameter (computed using \eq{e:MoL}),
the scales $r_0/a$ (where it is measured) and $t_0/a^2$ and 
the total statistics in molecular dynamics units. 
}\label{t:ens-tm}
\end{table}

%%%%%%%%%%%%%%
\begin{table}[t]
  \centering
  \begin{tabular}{ccc}
    \toprule
    $\beta$ & $L_1/a$ & $a$ [$\fm$] \\
    \midrule
    $5.30$ & 6.195(51) & 0.066 \\
    $5.50$ & 8.280(80) & 0.049 \\
    $5.60$ & 9.569(99) & $\approx$0.042 \\
    $5.70$ & 11.07(17) & $\approx$0.036 \\
    $5.88$ & 14.30(24) & $\approx$0.028 \\
    $6.00$ & 17.27(70) & $\approx$0.023 \\
\bottomrule
\end{tabular}
  \caption{The values of the scale $L_1/a$ used in our simulations and the
    corresponding lattice spacings.}
 \label{t:scale-L1}
\end{table}
%%%%%%%%%%%%%%
In \tab{t:scale-L1} we list the values of the hadronic scale $L_1/a$
\cite{hqet:paramnf2,alpha:lambdanf2}.
At $\beta=5.3,~5.5$ they are taken from Table 7 of \cite{alpha:lambdanf2}.
At the other $\beta$ values they are obtained
from a quadratic fit in $\beta$ of $\ln(L_1/a)$, where data for the latter
are taken from Table 13 of \cite{alpha:lambdanf2}.
The lattice spacing for $\beta>5.5$ (not covered by the simulations in \cite{alpha:lambdanf2})
can be inferred from the value $L_1=0.400(10)\fm$
determined in \cite{alpha:lambdanf2}.

\subsection{Mass corrections}

The data for a hadronic scale $\mscale$ such as
$r_0^{-1}$, $t_0^{-1/2}$ obtained from the simulations
with standard Wilson fermions are corrected for 
small mismatches of the values $M/\Lambda$ compared to the target values
$M_\mathrm{t}/\Lambda$ given in \eq{e:targetM},
see \tab{t:ens-Wilson}. This
is done by fitting the $\beta=5.7$ data to the form
\begin{equation}\label{eq:massfit}
a\mscale(M) = s_1\times\left(M/\Lambda\right)^\alpha \,,
\end{equation}
with fit coefficients $s_1$ and $\alpha$.
This fit formula is motivated by \eq{e:theequ} taking the asymptotic expression
$P=(M/\Lamq)^{\eta_0}$.
For example for $\mscale=1/\sqrt{t_0}$ we get $\alpha=0.123(2)$ and
for $\mscale=1/r_0$ we get $\alpha=0.139(12)$
which are close to $\eta_0=0.121212$.
The corrected values $\mscale(M_\mathrm{t})$ are computed as
\begin{equation}\label{eq:masscorr}
\ln(a\mscale(M_\mathrm{t})) = \ln(a\mscale(M)) + \alpha\ln(M_\mathrm{t}/M) \,.
\end{equation}
Note that \eq{eq:masscorr} being a small correction is applied for all 
lattice spacings $a$.
Moreover the $\Lambda$ parameter drops out in \eq{eq:masscorr}. 
Since the main contribution to the error on $M/\Lambda$ 
comes from $\Lambda L_1$, it does not affect the mass corrections. 
In order to determine the final error of $a\mscale(M_\mathrm{t})$,
we propagate the error of the exponent $\alpha$ and linearly add its
contribution (for a conservative estimate) multiplied by a factor of two.

No corrections is needed for the hadronic scales from twisted mass simulations
since their parameters are tuned for the target mass values, see \sect{s:quark_masses}.

\vskip 0.3cm

\noindent

\bibliographystyle{JHEP} 
\bibliography{latticen,qcd,latticen_fk,qcd_fk,partphys}

\end{document}